\documentclass[%
 aip,
 sd,%
 amsmath,amssymb,
 reprint,%
]{revtex4-1}

\usepackage{graphicx}% Include f files
\usepackage{dcolumn}% Align table columns on decimal point
\usepackage{multirow}
\usepackage{bm}% bold math
\usepackage{epstopdf}
\usepackage{enumerate}
\usepackage{color}
\newcommand{\smc}[1]{\scriptstyle\mathcal{#1}}
\newcommand{\ssmc}[1]{\scriptscriptstyle\mathcal{#1}}

\begin{document}

\preprint{AIP/123-QED}

\title{Modeling the lowest-cost splitting of a herd of cows by optimizing a cost function}

\author{Kelum Gajamannage}
\email{kdgajamannage@wpi.edu.}
\affiliation{Department of Mathematics, Clarkson University, Potsdam, NY 13699, USA}
\affiliation{Department of Mathematical Sciences, Worcester Polytechnic Institute, Worcester,  MA 01609, USA}
\author{Erik M. Bollt}
\email{bolltem@clarkson.edu.}
\affiliation{Department of Mathematics, Clarkson University, Potsdam,  NY 13699, USA}
\author{Mason A. Porter}
\email{mason@math.ucla.edu}
\affiliation{Department of Mathematics, University of California, Los Angeles, Los Angeles, California 90095, USA}
\affiliation{Mathematical Institute, University of Oxford, Oxford OX2 6GG, UK}
\affiliation{CABDyN Complexity Centre, University of Oxford, Oxford, OX1 1HP, UK}
\author{Marian S. Dawkins}
\email{marian.dawkins@zoo.ox.ac.uk.}
\affiliation{%
Department of Zoology, University of Oxford, OX1 3PS, UK
}%

\date{\today}
\begin{abstract}

Animals live in groups to defend against predation and to obtain food. However, for some animals --- especially ones that spend long periods of time feeding --- there are costs if a group chooses to move on before their nutritional needs are satisfied. If the conflict between feeding and keeping up with a group becomes too large, it may be advantageous for some animals to split into subgroups of animals with similar nutritional needs. We model the costs and benefits of splitting in a herd of cows using a cost function that quantifies individual variation in hunger, desire to lie down, and predation risk. We model the costs associated with hunger and lying desire as the standard deviations of individuals within a group, and we model predation risk as an inverse exponential function of group size. We minimize the cost function over all plausible groups that can arise from a given herd and study the dynamics of group splitting. We examine how the cow dynamics and cost function depends on the parameters in the model and consider two biologically-motivated examples: (1) group switching and group fission in a herd of relatively homogeneous cows, and (2) a herd with an equal number of adult males (larger animals) and adult females (smaller animals).

\end{abstract}

                              
\maketitle

\begin{quotation}

Although animals gain many advantages --- such as protection from predators --- from living in groups, they also incur considerable costs. For grazing animals, such as cows and antelopes, these costs include having to balance their own nutritional needs to stay in one place to feed with the need to keep up with a group and stop grazing when the rest of the herd moves on. If the nutritional needs of different individuals are sufficiently disparate, this can lead to the splitting of a group so that those with similar needs to graze, lie, and ruminate remain together. If a group of animals becomes too small, however, this can increase the risk of predation, as small groups are less able than large groups to defend themselves against predators. In this paper, we describe a cost function (CF) that balances predation risk (based on group size) with different individual needs for feeding and lying down to infer the sizes at which group splitting occurs. We model variation in hunger and lying desire using the standard deviation of individuals within a group, and we model predation risk as an inverse exponential function of group size. In a series of examples, we optimize the CF for each individual in a group of animals and examine when groups of cows split into smaller groups.

\end{quotation}

\section{Introduction} \label{sec:level1}

Animals gain many advantages from grouping and synchronizing their behavior --- including greater vigilance, coordinated defense against predators, and increased ability to find and defend food sources.\cite{elgar1989predator, krause2002living} However, living in large groups also carries disadvantages, such as increased risk of disease and parasitism,\cite{brown1986ectoparasitism, altizer2003social} having food stolen,\cite{di2001social} and interference with movement.\cite{usherwood2011flying} A ``perfect'' synchronization requires animals to change their activities at a communal time rather than at individual ideal times, and this can be costly for individuals.
 
The balance between synchrony and risk of predation is complex,\cite{creel2005responses, hebblewhite2002effects} and one possible approach for examining such a balance is with a cost function with components from synchrony and risk. When a group of animals becomes very large, the cost incurred through synchrony tends to exceed that incurred through risk, as a significant number of individuals change their desired activities (like eating or lying) to conform with communal decisions. Because of the balance, a cost function (CF) with components from synchrony and risk of predation should have at least one optimum point, and one should expect animal groups to split if they are too large. However, an optimum group size is not necessarily a ``stable'' group size. Supposing animals join a group one by one, a stable group size is a size at which there is no further fission of groups or switching of animals between groups.\cite{sibly1983optimal} Even when a group is already at its optimum for existing individuals, extra individuals can still benefit from joining the group. At some point, however, the group can become sufficiently large that it splits into two groups, as this benefits its members more than the overloaded single group.\cite{sibly1983optimal} A stable group size is therefore likely to be consistently larger than an optimum group size.\cite{sibly1983optimal, sumpter2010collective} 

Sometimes grouping can be even more complicated, as individuals within a group differ in many ways that relate to their fitnesses. For example, males and females in a herd differ in their nutritional needs. However, although they can benefit from staying with a mixed-sex group, some individuals may have to interrupt valuable feeding or lying time to keep up with a herd when it moves. \cite{ruckstuhl1999synchronise, conradt2000activity} It can be costly for such individuals to synchronize their activities with others, as they are forced to switch between eating, lying down, or moving at a communal time rather than at a time that is optimal for them as individuals.\cite{dostalkova2010go}  Alternatively, a group may split into subgroups that consist of individuals with similar switching times (such as all males and all females, or juveniles and adults), and then the costs of synchronization are lower. \cite{ruckstuhl2002sexual, conradt2009group, dostalkova2010go, ramos2015collective, merkle2015follow} Such synchronization costs depend on the different activities of animals in a group, so some animals (e.g., baboons) break up into subgroups for foraging, particularly when food is scarce, and then come together into larger groups for sleeping.\cite{schreier2012ecology}

Social splitting between two categories (e.g., male--female or calves--adults) has been examined using an ordinary-differential-equation (ODE) model, whose performance was tested using data on mixed-sexual grouping in red deer.\cite{conradt2000activity} However, even for animals in the same category (e.g., males), activity synchronization can vary significantly, as it can depend on the age, body mass, and health of animals. Consequently, category-based splitting can lead to groups in which animals are still heterogeneous across many other categories. Splitting of animals in different categories can also be seasonal; for example, in nature, mixed-sexual social grouping does not occur during the mating season.\cite{ruckstuhl1998foraging}

Communal decisions in herds are made either despotically by a dominant animal (or dominant animals) or democratically by the majority of individuals in a group,\cite{conradt2003group, king2009leaders, kerth2010group} and the corresponding groups are called ``despotic groups'' and ``democratic groups'', respectively.  Modeling of synchronization costs has suggested that costs for despotic groups tend to be higher than than those for democratic groups.\cite{conradt2003group}

The rest of our paper is structured as follows.  In Sec.~\ref{sec:biological_modeling}, we discuss biological modeling principles and the construction of a CF, which encapsulates the demands of hunger and lying desire of cattle, for groups of cows to stay together or break apart. In Sec.~\ref{sec:evolution_and_cost}, we describe a method of determining demands for hunger and lying desire using a CF and an evolution scheme (ES) that describes the change in state (eating, standing, and lying down) of cows. In Sec.~\ref{sec:analysis}, we examine the dynamics of cows and study the cost function for various parameter values. In Sec.~\ref{sec:examples}, we present two examples: (1) group-switching dynamics of cows when a herd that consists of adults splits into a maximum of three groups, and (2) a scenario in which a herd that consists of an even mixture of males and females splits into a maximum of two groups. In Sec.~\ref{sec:conc_discu}, we discuss our results and present ideas for future work.

\section{Biological modeling principles}\label{sec:biological_modeling}

We consider the behavior of cows (\emph{Bos taurus}), which make many daily decisions about staying with or leaving a herd. Cows have a two-stage feeding process that involves first grazing (standing up) and then ruminating (predominantly lying down). Together, lying down and ruminating can occupy up to 65\% of a cow's day. \cite{krohn1993behaviour, kilgour2012pursuit} Both grazing and lying (including ruminating) are essential for successful digestion of grass, \cite{schirmann2012rumination} but cows have to stop these actions if their herd decides to move to another area; this can occur 15--20 times a day. \cite{kilgour2012pursuit} Each individual cow has similar --- but not identical --- needs for lying and grazing, \cite{krohn1993behaviour, robert2011determination} so keeping up with a herd each time it moves can be considerably costly because of interrupted grazing or lying times. This cost can include reduced growth rate in young cattle \cite{mogensen1997association, nielsen1997resting} and physiological and behavioral symptoms of ``stress'' when a cow is deprived of adequate opportunities for lying down.\cite{fisher2002effects, munksgaard2005quantifying} 

Reference~\onlinecite{dostalkova2010go} considered costs from synchronization, as animals often need to change their behavior (e.g., staying in one place versus moving to another place) at a communal time rather than at their ideal time. In our work, we consider both a synchronization cost and a cost due to predation risk. We assume that large groups encounter a large synchronization cost and small groups increase the cost of predation risk.\cite{hebblewhite2002effects, creel2005responses} Therefore, an ``optimal" group size is neither too large nor too small. Moreover, we assume that the synchrony can vary within groups, so one set of cows can be eating while another set of cows is lying down or walking (in the neighborhood of others).

We construct a cost function (CF) based on the following four principal assumptions:
\begin{enumerate}[(i)]
\item Herds are fully democratic when cows take communal decisions, as this reduces cost.~\cite{conradt2003group}
\item Cows are free to switch between groups, which thus freely form or dissolve.\cite{albon1992cohort, raman1997factors, clutton1982red}
\item Fission of groups depends only on cows' hunger, lying desire, and predation risk.
\item The predation risk of a group is an exponential function of the inverse of the group size. 
\end{enumerate}
The decrease of predation risk with group size in assumption (iv) arises from the facts that having more animals in a group contributes to greater vigilance, \cite{elgar1989predator, krause2002living, pulliam1973advantages} a higher dilution effect, \cite{foster1981evidence, davies2012introduction}, and a higher confusion effect. \cite{neill1974experiments, parrish1993comparison} Consequently, a larger group size tends to result in a lower predation risk. Motivated by empirical studies in Refs.~\onlinecite{cresswell2011predicting, pulliam1973advantages, elgar1981flocking, krakauer1995groups}, which described an inverse exponential relationship between group size and predation risk, we use such a relationship between group size and predation risk of cows in assumption (iv).

We model the CF, which we denote by $C$ in Sec.~\ref{sec:cost_function}, as a convex combination of costs from hunger ($h$), desire to lie down ($f$), and predation risk ($r$). We thus write 
\begin{equation} \label{costfunction}
	C=\lambda h+\mu f+(1-\lambda-\mu) r\,,
\end{equation}
where $\lambda, \mu \in [0, 1]$ are parameters. In Eq.~\eqref{costfunction}, ``hunger" refers to the grazing demand of cows in a group, and ``lying desire" is their demand to lie down. We compute their hunger and lying desire at each time step using a previously-introduced evolution scheme (ES)~\cite{sun2011mathematical} for cows to change their state (where eating, standing, and lying down are the three possible states), and we quantify synchronization cost based on cows' hunger and desire to lie down. We assess the cost from hunger (respectively, lying desire) as the mean over all groups of the standard deviation of hunger (respectively, lying desire) within each group, and we model the cost from risk as a function of the group size. During each time step, we minimize the CF over all groups that one can construct from a given herd, where we specify a maximum number of groups, and determine the lowest splitting cost. We determine the optimum group sizes using the minimum of the CF, as it rewards groups with homogeneous demands for hunger and lying desire. This construction enforces perfect synchronization of activity within each group. Our modeling framework is very flexible, and we can consider more general situations by considering different CFs, measuring synchrony in different ways, and other generalizations.

\section{Temporal evolution and modeling group splitting}\label{sec:evolution_and_cost}

As in Ref.~\onlinecite{sun2011mathematical}, when considering a herd, we simulate cows' hunger (i.e., desire to eat) and lying desire (i.e., desire to lie down) and change of states between eating, lying down, and standing. We then present a CF and optimize it to determine the lowest-cost splitting of the herd.

\subsection{Temporal evolution and change of states of cows}\label{sec:temporal_evolution}

Cows interact with each other through the ES, which thereby helps provide some understanding of their cooperative activities. We augment the ES in Ref.~\onlinecite{sun2011mathematical} by formulating it as an iterative scheme that we combine with our CF. In this model, each individual cow is a piecewise-smooth dynamical system, and a cow switches between three discrete states: eating ($\smc{E}$), lying down ($\smc{R}$), and standing ($\smc{S}$). There are also continuous variables, $x\in[0,1]$ and $y\in[0,1]$, that, respectively, represent cows' desire to eat and desire to lie down. The dynamics of a single cow are given by the following set of differential equations:
\begin{equation}\label{eqn:single_cow}
\begin{split}
\text{Eating state ($\smc{E}$): } 
\begin{cases}
    \dot x=-\xi^{''}x\,,\\ 
    \dot y=\zeta^{'}y\,,\\
  \end{cases}\\
\text{Lying-down state ($\smc{R}$): } 
\begin{cases}
    \dot x=\xi^{'}x\,,\\ 
    \dot y=-\zeta^{''}y\,,\\
  \end{cases}\\
\text{Standing state ($\smc{S}$): } 
\begin{cases}
    \dot x=\xi^{'}x\,,\\ 
    \dot y=\zeta^{'}y\,, \\
  \end{cases}
\end{split}
\end{equation}
where
\begin{equation*}
	\begin{split}
\xi'_i &\ \text{is the rate of increase of hunger\,,}\\
\xi''_i &\ \text{is the decay rate of hunger\,,}\\
\zeta'_i &\ \text{is the rate of increase of desire to lie down\,, and}\\
\zeta''_i &\ \text{is the decay rate of desire to lie down}
	\end{split}
\end{equation*}
of the $i$th cow. The parameters $\xi'_i$, $\xi''_i$, $\zeta'_i$, and $\zeta''_i$ are all positive. These parameters can be different for different cows. If two cows have similar parameter values, we expect them to exhibit similar dynamics. Based on the hypothesis that it is good for cows to eat when other cows are eating and to lie down when other cows are lying down, one can extend the ``single-cow model'' in Eq.~(\ref{eqn:single_cow}) into a coupled dynamical system by allowing the individual cows to interact, and we use a time-dependent adjacency matrix to encode which cows are interacting with each other (see Sec.~\ref{sec:cost_evolution}). In Eq.~\eqref{eqn:diff_equation} below, we express how coupling influences the dynamics of cows.

As we mentioned previously, we modify the coupled system in Ref.~\onlinecite{sun2011mathematical} to produce an iterative scheme. To simplify our exposition (though at the cost of some technical correctness in the context of animal behavior), we sometimes use the terms ``lying desire'' to represent ``desire to lie down'' and ``hunger'' to represent ``desire to eat.'' Because we study the dynamics of the cows at each instant when the state variable changes from one state to another, we only record $x$ and $y$ for the cows at these instants. Thus, for $t \in \{1, \dots, T-1\}$ and $i \in \{1, \dots, n\}$, the discrete-time variables $x^{(t)}_i \in [0,1]$ and $y^{(t)}_i \in [0,1]$, respectively, represent the level of hunger and desire to lie down of the $i$th cow when the discrete-time state variable $\theta^{(t)}_i$ changes at time $t$. The variable $\theta^{(t)}_i$ represents the new state of cow $i$ at time $t$; it can be eating ($\smc{E}$), lying down ($\smc{R}$), or standing ($\smc{S}$). 

As one can see from the paragraph above, the $i$th cow is described by three variables: $\theta^{(t)}_i$, $x^{(t)}_i$, and $y^{(t)}_i$. For times $t \in \{1, \dots, T-1\}$ and cows $i \in \{1, \dots, n\}$, the time-dependent coupling of cows is given by the differential equations
\begin{equation}
	\begin{split}\label{eqn:diff_equation}
\dot x^{(t+1)}_i=\left[\alpha_i\big(\theta^{(t)}_i\big)+\frac{\sigma_x}{d^{(t)}_i}\sum^n_{j=1} a^{(t)}_{ij} \chi_{\ssmc{E}}\big(\theta^{(t)}_j\big)\right]x^{(t)}_i\,, \\
\dot y^{(t+1)}_i=\left[\beta_i\big(\theta^{(t)}_i\big)+\frac{\sigma_y}{d^{(t)}_i}\sum^n_{j=1} a^{(t)}_{ij} \chi_{\ssmc{R}}\big(\theta^{(t)}_j\big)\right]y^{(t)}_i\,,
	\end{split}
\end{equation}
where
\begin{equation}
	\begin{split}
\alpha_i\big(\theta^{(t)}_i\big) &:= -\xi^{''}_i\chi_{\ssmc{E}}\big(\theta^{(t)}_i\big) +\xi^{'}_i\chi_{\ssmc{R}}\big(\theta^{(t)}_i\big) 
+\xi^{'}_i\chi_{\ssmc{S}}\big(\theta^{(t)}_i\big)\,, \\
\beta_i\big(\theta^{(t)}_i\big) &:= \zeta^{'}_i\chi_{\ssmc{E}}\big(\theta^{(t)}_i\big) -\zeta^{''}_i\chi_{\ssmc{R}}\big(\theta^{(t)}_i\big) 
+\zeta^{'}_i\chi_{\ssmc{S}}\big(\theta^{(t)}_i\big)\,,
	\end{split}
\end{equation}
with
\begin{equation}
\chi_{\psi}\big(\theta^{(t)}_i\big)=\left\{\def\arraystretch{1.2}%
\begin{array}{@{}c@{\quad}l@{}}
    1\,, & \text{$\theta^{(t)}_i=\psi$\,,}\\
    0\,, & \text{otherwise\,.}\\
  \end{array}\right.
\end{equation}

The time-dependent adjacency matrix $A^{(t)}=\left[a^{(t)}_{ij}\right]_{n\times n}$ represents a network of cows at time $t$. Its components are
\begin{equation}
	a^{(t)}_{ij}=\left\{\def\arraystretch{1.2}%
\begin{array}{@{}c@{\quad}l@{}}
    1\,, & \text{if the $i$th cow interacts with}\\
	& \text{the $j$th cow at time $t$\,,}\\
    0\,, & \text{otherwise\,.}\\
  \end{array}\right.
\end{equation}
Thus, $d^{(t)}_i=\sum^n_{j=1}a^{(t)}_{ij}$ is the degree (i.e., the number of other cows with which it interacts) of cow $i$. We will discuss such interactions in terms of cow groupings in Sec.~\ref{sec:cost_evolution}. The nonnegative parameters $\sigma_x$ and $\sigma_y$, respectively, represent coupling strengths with respect to hunger and desire to lie down.

The switching condition of the state variable $\theta^{(t)}_i$ of the $i$th cow at time step $t$ is
\begin{equation}\label{eqn:state_change}
	\theta^{(t+1)}_i \rightarrow \left\{\def\arraystretch{1.2}%
\begin{array}{@{}c@{\quad}l@{}}
    	\smc{E}, & \text{if $\theta^{(t)}_i\in\left\{\smc{R}\,, \smc{S}\right\}$ and $x^{(t)}_i=1$\,,}\\	
   	\smc{R}, & \text{if $\theta^{(t)}_i\in\left\{\smc{E}\,, \smc{S}\right\}$ and $x^{(t)}_i<1$, $y^{(t)}_i=1$\,,}\\
	\smc{S}, & \text{if $\theta^{(t)}_i\in\left\{\smc{E}\,, \smc{R}\right\}$ and $x^{(t)}_i<1$, $y^{(t)}_i=\delta$}\\
	      		& \hspace{65pt} \text{(or $x^{(t)}_i=\delta$, $y^{(t)}_i<1$)\,,}\\
\end{array}\right.
\end{equation}
where we use the parameter $\delta \in (0,1)$ to exclude the point $\big(x^{(t)}_i, y^{(t)}_i\big)=(0, 0)$ from the variable domain (because it creates degenerate solutions). 

\begin{figure}[hpt]  	  
	\centering   
        	\includegraphics[width=.48\textwidth]{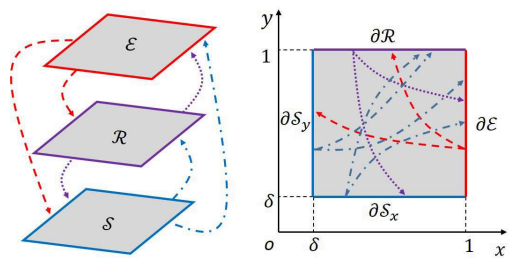}	
        	\caption{Schematic of the switching dynamics of a cow. The left panel is a new version of the right panel of Fig.~1 in Ref.~\onlinecite{sun2011mathematical}, and the right panel is an integrated version of the four panels from Fig.~2 of Ref.~\onlinecite{sun2011mathematical}. The left panel shows three states ($\{\mathcal{E}, \mathcal{R}, \mathcal{S}\}$, where $\mathcal{S}=\mathcal{S}_x \cup \mathcal{S}_y$) and the possibilities for switching between states. The edges of the square in the right panel represent boundaries of the domain of the continuous variables $x$ and $y$ [see Eq.~(\ref{eqn:boundaries})]. We are interested only in the discrete dynamics of cows; they are given by $x^{(t)}_i$, $y^{(t)}_i$, and $\theta^{(t)}_i$ on the boundaries. The arrows represent all possible state switches of a cow. A given style and color of arrows in the left and right panels indicates the same type of switch.}
	\label{fig:state_change}	
\end{figure}

We study the dynamics of cows at discrete times, so we examine a Poincar\'e section that we construct (using ideas from Ref.~\onlinecite{gouze2002class}) by considering switches between different states. We can thereby study the dynamics given by Eqs.~\eqref{eqn:diff_equation} and \eqref{eqn:state_change}. See the schematic in Fig.~\ref{fig:state_change}. The boundaries of this Poincar\'e section are
\begin{equation}\label{eqn:boundaries}
	\begin{split}
\partial \mathcal{E}=\big\{\big(x, y, \theta\big)\big\vert x=1\,, \delta\le y\le1\,, \theta=\mathcal{E}\big\}\,,\\
\partial \mathcal{R}=\big\{\big(x, y, \theta\big)\big\vert \delta \le x<1\,, y=1\,, \theta=\mathcal{R}\big\}\,,\\
\partial \mathcal{S}_x=\big\{\big(x, y, \theta\big)\big\vert \delta < x<1\,, y=\delta, \theta=\mathcal{S}\big\}\,,\\
\partial \mathcal{S}_y=\big\{\big(x, y, \theta\big)\big\vert x=\delta, \delta\le y<1\,, \theta=\mathcal{S}\big\}\,.\\
	\end{split}
\end{equation}
These four boundaries arise from the switching conditions in \eqref{eqn:state_change}; the first pair of conditions yields the first two boundaries, and the second pair yields the last two boundaries. At time $t$, the variables $x^{(t)}_i$, $y^{(t)}_i$, and $\theta^{(t)}_i$ of the $i$th cow are represented by one of the boundaries, and then the cow switches to another boundary in the subsequent time step according to the switching condition in Eq.~\eqref{eqn:state_change}.

We solve the dynamical system in Eq.~(\ref{eqn:diff_equation}) for $n$ cows in $T$ time steps together with the switching condition in Eq.~(\ref{eqn:state_change}). The solution gives the discrete dynamics of the $i$th cow in terms of $x^{(t)}_i$, $y^{(t)}_i$, and $\theta^{(t)}_i$ at each time step $t$. We show the derivation of these solutions in Appendix \ref{sec:appendix} as an iterative scheme. As one can see in the left panel of Fig.~\ref{fig:state_change}, at time step $t$, each cow is in one of three states ($\mathcal{E}$, $\mathcal{R}$, or $\mathcal{S}$) at the start of the time step, and it switches to one of the other two states, where it starts the $(t+1)$th time step. The last two equations in Eq.~\eqref{eqn:boundaries} collectively explain the standing state, so both the lower and the left boundaries of the Poincar\'e section (see the right panel of Fig.~\ref{fig:state_change}) represent the standing state. Thus, in the Poincar\'e section, the starting point of each cow at a given time step is on one of four boundaries, and the end point at that time step is on a boundary that represents a new state (for which there are two possibilities). We present the corresponding iterative scheme of the solution in Table \ref{tab:iterative_scheme}, in which we use the following notation:
\begin{equation}\label{eqn:subs_diff}
	\begin{split}
\eta'_i &:=\xi'_i+\frac{\sigma_x}{d^{(t)}_i}\sum^n_{j=1} a^{(t)}_{ij} \chi^{(t)}_{\ssmc{E}}\left(\theta^{(t)}_j\right)\,,\\
\eta''_i &:=-\xi''_i+\frac{\sigma_x}{d^{(t)}_i}\sum^n_{j=1} a^{(t)}_{ij} \chi^{(t)}_{\ssmc{E}}\left(\theta^{(t)}_j\right)\,,\\
\gamma'_i &:=\zeta'_i+\frac{\sigma_y}{d^{(t)}_i}\sum^n_{j=1} a^{(t)}_{ij} \chi^{(t)}_{\ssmc{R}}\left(\theta^{(t)}_j\right)\,,\\
\gamma''_i &:=-\zeta''_i+\frac{\sigma_y}{d^{(t)}_i}\sum^n_{j=1} a^{(t)}_{ij} \chi^{(t)}_{\ssmc{R}}\left(\theta^{(t)}_j\right)\,,\\
	\end{split}
\end{equation}
where $i \in \{1, \dots, n\}$ and $t \in \{1, \dots, T\}$.

\begin{table*}[htp]
\caption{Iterative scheme for temporal evolution of cow dynamics that we obtain from solving the dynamical system in Eq.~\eqref{eqn:diff_equation} with the switching condition in Eq.~\eqref{eqn:state_change}. We show the derivation of these solutions in Appendix \ref{sec:appendix}. For the $i$th cow at time step $t$, one of cases 1, 2, 3, and 4 in this table represents the boundary of the Poincar\'e section (see Eq.~\eqref{eqn:boundaries}) associated with the cow at the beginning of that time step. For each of the four situations, subcases ``a'' and ``b'' represent the new state of the cow at the end of time step $t$. We illustrate all eight possible combinations in the Poincar\'e section in Fig.~\ref{fig:sh_state_change1}.}

\begin{ruledtabular}
\begin{tabular}{l l l l}
&\\
case 1: & \multicolumn{3}{l}{If $x^{(t)}_i=1$, $\delta \le y^{(t)}_i\le1$, and $\theta^{(t)}_i=\smc{E}$}\\

[Eqs.~\eqref{eqn:d1} and ~\eqref{eqn:ds1}] & subcase a:  if $y^{(t)}_i\ge \delta^{\frac{\gamma'_i}{\eta''_i}}$, 
& then& $x^{(t+1)}_i=\left[y^{(t)}_i\right]^{\frac{\eta''_i}{\gamma'_i}}$, $y^{(t+1)}_i=1$, and  $\theta^{(t+1)}_i=\smc{R}$\\

& subcase b:  if $y^{(t)}_i < \delta^{\frac{\gamma'_i}{\eta''_i}}$, 
& then& $x^{(t+1)}_i=\delta$,  $y^{(t+1)}_i=\delta^{-\frac{\gamma'_i}{\eta''_i}}y^{(t)}_i$, and $\theta^{(t+1)}_i=\smc{S}$\\
& & & \\

\hline
&\\
case 2: & \multicolumn{3}{l}{If $\delta \le x^{(t)}_i<1$, $y^{(t)}_i=1$, and $\theta^{(t)}_i=\smc{R}$} \\

[Eqs.~\eqref{eqn:d2} and ~\eqref{eqn:ds2}] & subcase a:  if $x^{(t)}_i\ge \delta^{\frac{\eta'_i}{\gamma''_i}}$, 
& then& $x^{(t+1)}_i=1$, $y^{(t+1)}_i=\left[x^{(t)}_i\right]^{\frac{\gamma''_i}{\eta'_i}}$, and $\theta^{(t+1)}_i=\smc{E}$\\

& subcase b:  if $x^{(t)}_i < \delta^{\frac{\eta'_i}{\gamma''_i}}$, 
& then& $x^{(t+1)}_i=\delta^{-\frac{\eta'_i}{\gamma''_i}}x^{(t)}_i$, $y^{(t+1)}_i=\delta$, and $\theta^{(t+1)}_i=\smc{S}$\\
& & & \\

\hline
&\\
case 3: & \multicolumn{3}{l}{If $x^{(t)}_i=\delta$, $\delta \le y^{(t)}_i<1$, and $\theta^{(t)}_i=\smc{S}$} \\

[Eqs.~\eqref{eqn:d3} and ~\eqref{eqn:ds3}] & subcase a:  if $y^{(t)}_i\le \delta^{\frac{\gamma'_i}{\eta'_i}}$, 
& then& $x^{(t+1)}_i=1$, $y^{(t+1)}_i=\delta^{-\frac{\gamma'_i}{\eta'_i}}y^{(t)}_i$, and  $\theta^{(t+1)}_i=\smc{E}$\\

& subcase b:  if $y^{(t)}_i > \delta^{\frac{\gamma'_i}{\eta'_i}}$, 
& then& $x^{(t+1)}_i=\left[y^{(t)}_i\right]^{-\frac{\eta'_i}{\gamma'_i}}\delta$, $y^{(t+1)}_i=1$, and  $\theta^{(t+1)}_i=\smc{R}$ \\
& & & \\

\hline
&\\
case 4: & \multicolumn{3}{l}{If $\delta<x^{(t)}_i<1$, $y^{(t)}_i=\delta$, and $\theta^{(t)}_i=\smc{S}$} \\

[Eqs.~\eqref{eqn:d3} and ~\eqref{eqn:ds4}] & subcase a:  if $x^{(t)}_i\ge \delta^{\frac{\eta'_i}{\gamma'_i}}$, 
& then& $x^{(t+1)}_i=1$, $y^{(t+1)}_i=\left[x^{(t)}_i\right]^{-\frac{\gamma'_i}{\eta'_i}}\delta$, and $\theta^{(t+1)}_i=\smc{E}$\\

& subcase b:  if $x^{(t)}_i < \delta^{\frac{\eta'_i}{\gamma'_i}}$, 
& then& $x^{(t+1)}_i=\delta^{-\frac{\eta'_i}{\gamma'_i}}x^{(t)}_i$, $y^{(t+1)}_i=1$ , and  $\theta^{(t+1)}_i=\smc{R}$\\
& & & \\
\end{tabular}\label{tab:iterative_scheme}
\end{ruledtabular}
\end{table*}

\begin{figure}[hpt]  
	\vspace{10pt}   
	\centering   
        	\includegraphics[width=.33\textwidth]{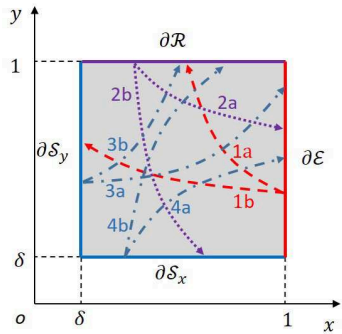}	
        	\caption{All possible state switches of a cow in a single time step.\cite{sun2011mathematical} This figure is an integrated version of the four panels from Fig.~2 of Ref.~\onlinecite{sun2011mathematical}. The states are $\{\mathcal{E}, \mathcal{R}, \mathcal{S}\}$ (eating, lying down, and standing), where $\mathcal{S}=\mathcal{S}_x \cup \mathcal{S}_y$. We show all possible state transitions in Table \ref{tab:iterative_scheme}. For example, ``1a'' refers to ``subcase a'' of ``case 1'' in the table. We use the same style and color of arrows as in Fig.~\ref{fig:state_change}.}
	\label{fig:sh_state_change1}	
\end{figure}

\subsection{Cost function (CF) determining the splitting of herds}\label{sec:cost_function}

In this section, we determine the lowest-cost grouping of cows by minimizing a CF. This gives the total number of groups and the number of cows in each group. We suppose that a herd of cows splits into a maximum of $L$ distinct groups $N^{(t)}_1, \dots, N^{(t)}_L$, with $|N^{(t)}_l|=n^{(t)}_l>0$ cows in the $l$th group (where $l \in \{1, \dots, L\}$), at each time step $t\in T$. If the herd splits into $L_1<L$ groups at some time step $t_1$, we set $|N^{(t_1)}_l|=0$ for $l \in \{L_1+1, \dots, L\}$.

Our CF is the sum of two components: a synchronization component and a risk component. The synchronization component (SC) models the cost due to variation in the lying desire and/or hunger of cows, and the risk component (RC) models the cost from predation risk. 

\subsubsection{Synchronization component}\label{sec:synchronization_component}

Recall from Sec.~\ref{sec:temporal_evolution} that cow $i$'s hunger is $x^{(t)}_i\in[0, 1]$ and lying desire is $y^{(t)}_i\in[0, 1]$. We re-index the variables $x^{(t)}_i$ as $x^{(t)}_{k,l}$ and $y^{(t)}_i$ as $y^{(t)}_{k,l}$, respectively, to denote the hunger and lying desire of the $k$th cow in the $l$th group at the $t$th time step. Because hunger and lying desire are separate motivations in cows, we compute the two groupings independently, so that cows are optimally homogeneous with respect to hunger (case I) or optimally homogeneous with respect to lying desire (case II). Of these two groupings, we then select the one with the lower synchronization cost.

\textbf{Case I}: We sort cows according to increasing hunger, and we place the first $n^{(t)}_1$ cows into group $N^{(t)}_1$, the next $n^{(t)}_2$ cows into group $N^{(t)}_2$, and so on. In each group, the synchronization cost from hunger represents the heterogeneity of hunger within the group. As a simple way to quantify this cost, we use the mean of the standard deviations of cows' hunger within the groups and thus calculate 
\begin{equation}\label{eqn:hun}
	h_1^{(t)}=\frac{1}{L}\sum_{l=1}^{L}\sqrt{\frac{\sum_{k=1}^{n^{(t)}_l}\left(x^{(t)}_{k,l}-\sum_{k=1}^{n^{(t)}_l}x^{(t)}_{k,l}\Big{/}n^{(t)}_l\right)^2}{n^{(t)}_l}}
\end{equation}
to assess the SC due to hunger. Similarly, we quantify the heterogeneousness of groups with respect to lying desire as the mean of the standard deviations of cows' lying desire in groups by calculating
\begin{equation}\label{eqn:fat}
	f_1^{(t)}=\frac{1}{L}\sum_{l=1}^{L}\sqrt{\frac{\sum_{k=1}^{n^{(t)}_l}\left(y^{(t)}_{k,l}-\sum_{k=1}^{n^{(t)}_l}y^{(t)}_{k,l}\Big{/}n_l\right)^2}{n^{(t)}_l}}\,.
\end{equation}
 
 \textbf{Case II}: Similar to case I, we sort cows according to increasing lying desire and place them into groups $N^{(t)}_1, \dots, N^{(t)}_L$. We again compute hunger and lying desire from the means of the standard deviations within the groups, and we denote them by $h_2^{(t)}$ and $f_2^{(t)}$, respectively. 

From the cow groups that we find in cases I and II, we choose the grouping that yields the minimum SC. The hunger $h^{(t)}$ and the lying desire $f^{(t)}$ of the cow herd at time $t$ are thus
\begin{equation}
	h^{(t)}=\min\left\{h_1^{(t)}, h_2^{(t)}\right\}\,, \quad \ f^{(t)}=\min\left\{f_1^{(t)}, f_2^{(t)}\right\}\,.
\end{equation}

\subsubsection{Risk component}\label{sec:risk_component}

\begin{figure}[hpt]     
	\centering   
        	\includegraphics[width=.3\textwidth]{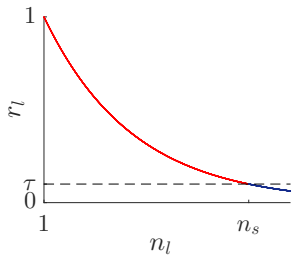}
        	\caption{We model predation risk $r_l$ as an inverse exponential function of group size $n_l$. We use $\tau$ to denote the safety level and $n_s$ to denote the safe size. The red segment of the curve signifies a regime with an unsafe level of risk, and the blue segment signifies a regime with a safe level of risk.}
	\label{fig:risk_comp}
\end{figure}

Unlike hunger and lying desire, a herd's predation risk is independent of individuals' states and depends only on the group size. Group size is related inversely to the risk of being attacked by predators\cite{elgar1989predator, krause2002living, pulliam1973advantages, foster1981evidence, davies2012introduction, neill1974experiments, parrish1993comparison}, and we model the predation risk $r_l\in (0, 1]$ of the $l$th group (which has size $n_l>0$) as an inverse exponential function of group size: \cite{cresswell2011predicting, pulliam1973advantages, elgar1981flocking, krakauer1995groups}
\begin{equation}\label{eqn:risk1}
	r_l=e^{-\left(1-n_l\right)/c}\,,
\end{equation} 
where $c$ is a constant. We assume that the predation risk is small when a group has sufficiently many cows, and we use this condition to compute the constant $c$. We denote this sufficient group size (the so-called ``safe size") by $n_s$, and we denote the small risk to which the risk function converges (the so-called ``safety level") at this size by $\tau$. The constant $c$ is thus $-(1-n_s)/\ln \tau$, so the RC of the group is
\begin{equation}\label{eqn:risk11}
	r_l=\tau^{\left(\frac{1-n_l}{1-n_s}\right)}\,.
\end{equation}
In Fig.~\ref{fig:risk_comp}, we show the relationship given by the Eq.~(\ref{eqn:risk11}) for a group with safety level $\tau$ and safe size $n_s$.

We compute the predation risk of each group, and we treat its mean
\begin{equation}\label{eqn:risk2}
	r=\frac{1}{L}\sum_{l=1}^{L}\tau^{\left(\frac{1-n_l}{1-n_s}\right)}\,,
\end{equation}
as the risk of the herd.

\begin{figure}[hpt]     
	\centering   
        	\includegraphics[width=.3\textwidth]{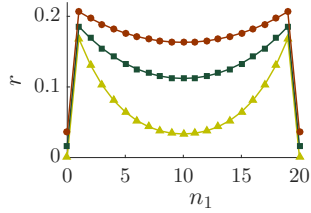}
        	\caption{Risk of splitting a herd of $n=20$ cows into two groups of sizes $n_1$ and $n_2=20-n_1$, where the safety level is $\tau= 1/30$ and the safe size is $n_s= 10$ (yellow triangles), $n_s = 20$ (green squares), and $n_s = 30$ (brown disks).}
	\label{fig:risk_twog}
\end{figure}

In real situations, the safe size and safety level depend on the environment in which a herd lives. If the environment is either dense with predators or vulnerable to predation, the safe size should be comparatively large to achieve a significant safety level. As an example, we use Eq.~(\ref{eqn:risk2}) and compute the risk of splitting a herd of $n=20$ cows into two groups with a safety level of $\tau= 1/30$ and safe sizes of $n_s= 10$, $n_s = 20$, and $n_s = 30$ (see Fig.~\ref{fig:risk_twog}). We thereby illustrate that large safety sizes model riskier situations for a herd than small safety sizes, independently of how the herd splits. For all safety sizes, we achieve the lowest cost when the herd remains intact (i.e., no splitting), because larger group sizes entail safer herds. We achieve the second-lowest cost when the herd splits into equal-sized groups.

\vspace{15pt}

\subsubsection{Cost function}\label{sec:cost_function}

We formulate the CF as a convex combination of the costs from hunger, lying desire, and risk of predation:
\begin{equation} \label{cost2}
	C^{(t)}(n^{(t)}_1, \dots, n^{(t)}_L)=\lambda h^{(t)}+\mu f^{(t)}+(1-\lambda-\mu)r\,,
\end{equation}
where $\lambda, \mu \in[0, 1]$. For a given herd, which we denote by the set $N$, and a maximum number $L$ of groups into which it can split, we minimize \eqref{cost2} over all plausible groups that can be created, and we thereby determine the lowest-cost splitting.

\subsection{Cost function and temporal evolution}\label{sec:cost_evolution}

\begin{figure*}[hpt]     
	\centering   
	\includegraphics[width=.7\textwidth]{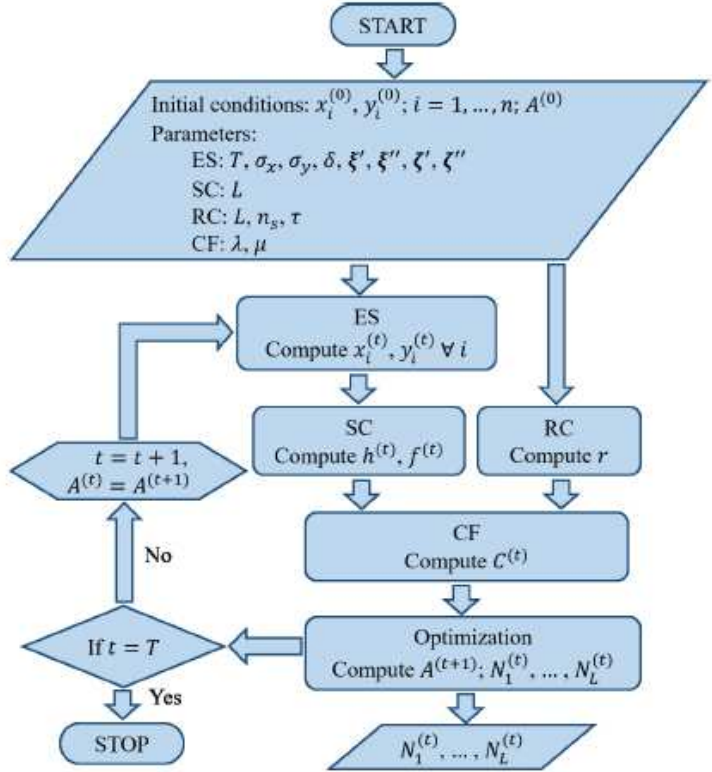}
        	\caption{Flow chart for our model. The inputs are (1) initial conditions for the variables for hunger and lying desire and (2) values for the parameters associated with ES, SC, RC, and CF. We explain these parameters in Secs.~\ref{sec:temporal_evolution}, \ref{sec:synchronization_component}, \ref{sec:risk_component}, and \ref{sec:cost_function}, respectively. At each time step, we adjust the adjacency matrix, which encodes which cows interact with each other, using the new grouping information that we obtain by optimizing the CF. At each time step, our model outputs the groups of animals that correspond to the lowest-cost splitting, and it terminates upon reaching the stopping criterion (i.e., after a designated number of time steps).}
	\label{fig:algorithm}
\end{figure*}

We examine the CF simultaneously with the temporal ES for times $t \in \{1, \dots, T\}$. At each time step, we update the adjacency matrix $A^{(t)}=\left[a^{(t)}_{i,j}\right]_{n\times n}$ in the scheme so that it agrees with the best grouping provided by the optimization of the CF in the previous time step. That is, 
\begin{equation}\label{eqn:state_switching}
	a^{(t+1)}_{i,j} = \left\{\def\arraystretch{1.2}%
\begin{array}{@{}c@{\quad}l@{}}
    	1\,, & \text{if $i, j \in N^{(t)}_l$\,,}\\	
   	0\,, & \text{otherwise\,,}\\	
\end{array}\right.\, \qquad l\in\{1, \dots, L\}\,.
\end{equation}
This adjacency matrix, which encodes the network architecture of a herd, is an input in Ref.~\onlinecite{sun2011mathematical}. However, in this paper, we update the adjacency matrix at each time step based on an optimum grouping. At each time step, optimizing the CF outputs a lowest-cost grouping until we reach a stopping criterion, which we take to be the maximum time $T$. In Fig.~\ref{fig:algorithm}, we show a flow chart of this process.

\vspace{10pt}

\section{Exploration of parameter space}\label{sec:analysis}

We now explore how the effects of the parameters in our model. We first examine the dynamics of cows for different coupling strengths, and we then study the CF for different values of the parameters $\sigma_x$, $\sigma_y$, $\tau$, and $n_s$ and different coupling strengths. 

\subsection{Cow dynamics}\label{sec:dynamics}

We explore the dynamics of cows with respect to coupling strength by examining hunger and lying desire on the boundary of a Poincar\'e section. We then compute the mean group size of a herd for different safety levels. 

We undertake these explorations using one herd of $n=12$ cows that splits into a maximum of three groups ($L=3$). We simulate hunger and lying desire using the ES (see Sec.~\ref{sec:temporal_evolution}) followed by computing the CF \eqref{cost2} and optimizing it to determine a lowest-cost grouping at each time $t\in T$. As we will discuss shortly, we draw some of the initial conditions and parameter values from probability distributions.

In the ES, we set the initial states of cows to be $\theta^{(0)}_i\in\mathbb{U}\left\{\smc{E}, \smc{R}, \smc{S}\right\}$ for $i \in \{1, \dots, n\}$, where $\mathbb{U}$ denotes a uniform probability distribution over the set in its argument.  We add noise sampled from a uniform distribution into the initial conditions and parameters, as it is the simplest type of noise to consider. We determine the initial conditions $x^{(0)}_i$ and $y^{(0)}_i$ as follows:
\begin{equation}\label{eqn:ini_states_ex1}
	\left\{\def\arraystretch{1.2}%
\begin{array}{@{}c@{\quad}l@{}}
    x^{(0)}_i=1 \text{ and } y^{(0)}_i\in\mathbb{U}[\delta, 1]\,, & \text{ if } \theta^{(0)}_i=\smc{E}\,,\\
    x^{(0)}_i\in\mathbb{U}[\delta, 1) \text{ and } y^{(0)}_i=1\,, & \text{ if } \theta^{(0)}_i=\smc{R}\,,\\
	\left\{\def\arraystretch{1.2}%
	\begin{array}{@{}c@{\quad}l@{}}
    	x^{(0)}_i=\delta \text{ and } y^{(0)}_i\in\mathbb{U}[\delta, 1)\,,\\
	\text{or}\\
    	x^{(0)}_i\in\mathbb{U}(\delta, 1) \text{ and } y^{(0)}_i=\delta\,, \\
     	 \end{array}\right.
     & \text{ if } \theta^{(0)}_i=\smc{S}\,.\\
  \end{array}\right.
\end{equation}
For $\theta^{(0)}_i = \smc{S}$, each of the two subcases in Eq.~(\ref{eqn:ini_states_ex1}) has a 50\% chance of being the initial condition. We also make the following parameter choices for the ES: $\xi'_i\in\mathbb{U}[.0995, .1005]$, $\xi''_i\in\mathbb{U}[.1495, .1505]$, $\zeta'_i\in\mathbb{U}[.0495, .0505]$, $\zeta''_i\in\mathbb{U}[.1995, .2005]$, and $\delta=.25$.

\begin{figure}[!htp]
\centering
\vspace{20pt}
\includegraphics[width=.475\textwidth]{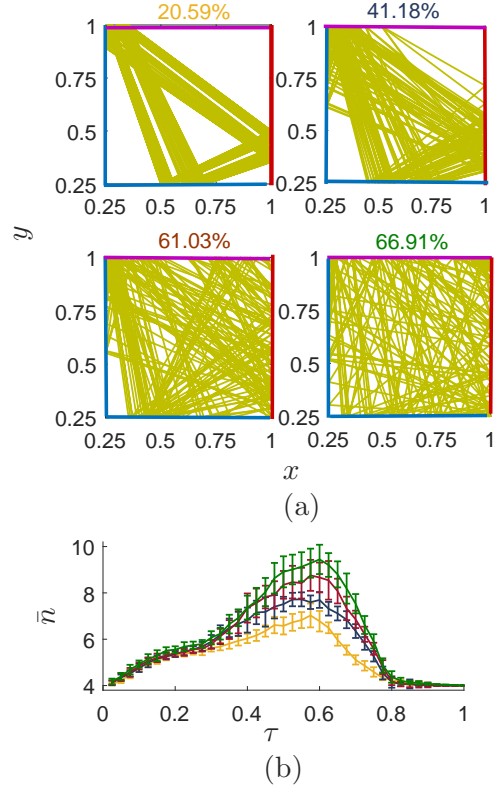}     
\caption{Herd dynamics for different coupling strengths for a herd of $n = 12$ cows. 
(a) We run the ES with $\xi'_i\in\mathbb{U}[.0995, .1005]$, $\xi''_i\in\mathbb{U}[.1495, .1505]$, $\zeta'_i\in\mathbb{U}[.0495, .0505]$, $\zeta''_i\in\mathbb{U}[.1995, .2005]$, $\delta=.25$, and $T=400$; an RC with $n_s=4$ and $\tau=.2$; an SC with $L=3$; and a CF with $\lambda=.33$ and $\mu=.33$. We generate orbits of hunger and lying desire of the first cow for (top left) $\sigma_x=\sigma_y=0$, (top right) $\sigma_x=\sigma_y=.05$, (bottom left) $\sigma_x=\sigma_y=.2$, and (bottom right) $\sigma_x=\sigma_y=.6$. The red, purple, and blue boundaries, respectively, represent the eating, lying, and standing states. The colors of the boundaries are the same as in Figs.~\ref{fig:state_change} and \ref{fig:sh_state_change1}. At the top of each Poincar\'e section, we show the percentage of the boundary that the orbit intersects; the color of the text matches the coupling strengths of the orbits to the coupling strengths of the plots in panel (b). (b) Means and standard deviations of the group sizes. We use the same parameter values for ES, SC, RC, and CF as in panel (a) and compute the mean group size $\bar{n}$ of a herd versus the safety level $\tau$ for coupling strengths $\sigma_x=\sigma_y=0$ (orange), $\sigma_x=\sigma_y=.05$ (blue), $\sigma_x=\sigma_y=.2$ (brown), and $\sigma_x=\sigma_y=.6$ (green) for $T = 400$ time steps.}
\label{fig:orbit}
\end{figure}

Cows are social animals, and their behavior is influenced by what other cows are doing.\cite{stoye2012synchronized} We model such interactions mathematically using the coupling parameters $\sigma_x$ and $\sigma_y$ in the ES. Using different values for these parameters in different simulations allows us to examine different biological scenarios, such as strongly interacting cows versus weakly interacting cows, and it can be helpful for understanding the dynamics of state changes of cows in these different scenarios. As an initial example, we let $\sigma_x=0$ and $\sigma_y=0$ (i.e., uncoupled cows) and run the ES for time $T=400$ to simulate hunger $x^{(t)}_i$ and lying desire $y^{(t)}_i$ for $i \in \{1, \dots, 12\}$ and $t \in \{1, \dots, 400\}$. We also set $n_s=4$ and $\tau=.2$ in the RC; $L=3$ in the SC; and $\lambda=.33$ and $\mu=.33$ in the CF. We then consider three other values of the coupling strengths: $(\sigma_x, \sigma_y) = (.05, .05)$, $(\sigma_x, \sigma_y) = (.2, .2)$, and $(\sigma_x, \sigma_y) = (.6, .6)$. In each case, we simulate $x^{(t)}_i$ and $y^{(t)}_i$ for $i \in \{1, \dots, 12\}$ and $t \in \{1, \dots, 400\}$. We determine $x_i(0)$ and $y_i(0)$ from Eq.~\eqref{eqn:ini_states_ex1} with initial states $\theta^{(0)}_i\in\mathbb{U}\left\{\smc{E}, \smc{R}, \smc{S}\right\}$ for $i \in \{1, \dots, 12\}$ a single time (as opposed to determining different values from the same distribution) and perform all four simulations with the same parameter choices (aside from coupling strengths). For each of the four cases above, we simulate one realization of the dynamics. In Fig.~\ref{fig:orbit}(a), we show the hunger and lying desire for cow $i = 1$. The uncoupled case is in the top-left panel, and the coupled cases are in the top-right panel ($(\sigma_x, \sigma_y) = (.05, .05)$), bottom-left panel ($(\sigma_x, \sigma_y) = (.2, .2)$), and bottom-right panel ($(\sigma_x, \sigma_y) = (.6, .6)$).
 
We observe similar dynamics for the other cows as that of the first cow [see Fig.~\ref{fig:orbit}(a)] when they do not interact with each other, but this is not the case when cows are allowed to interact with each other. 
In other words, $(\sigma_x, \sigma_y)= (0, 0)$ corresponds to modeling cows as independent oscillators, whereas the other cases correspond to coupled oscillators. To compare the dynamics in the four cases $(\sigma_x, \sigma_y) = (0, 0)$, $(\sigma_x, \sigma_y) = (.05, .05)$, $(\sigma_x, \sigma_y) = (.2, .2)$, and $(\sigma_x, \sigma_y) = (.6, .6)$, we measure the percentage of the length of the boundary of the Poincar\'e section that the orbit $(x, y)$ intersects in each case. To do this, we discretize each side of the boundary in the Poincar\'e sections into 75 intervals $[0.25, 0.26)$, $[0.26, 0.27)$, $\dots$, $[0.99, 1]$, and we then compute the percentage of the number of intervals that the orbit intersects. For $(\sigma_x, \sigma_y) = (0, 0)$, $(\sigma_x, \sigma_y) = (.05, .05)$, $(\sigma_x, \sigma_y) = (.2, .2)$, and $(\sigma_x, \sigma_y) = (.6, .6)$, these percentages are about $20.59\%$, $41.18\%$, $61.03\%$, and $66.91\%$, respectively. 

To examine the effect of different coupling strengths in a biological context, we compute the mean group size of the herd (which has $n=12$ cows) with respect to safety levels and coupling strengths. We first consider $\sigma_x=\sigma_y=0$ and set the initial states of the ES using $\theta^{(0)}_i\in\mathbb{U}\left\{\smc{E}, \smc{R}, \smc{S}\right\}$ for $i \in \{1, \dots, 12\}$ and then initial conditions using Eq.~\eqref{eqn:ini_states_ex1}. We use the parameter values $\xi'_i\in\mathbb{U}[.0995, .1005]$, $\xi''_i\in\mathbb{U}[.1495, .1505]$, $\zeta'_i\in\mathbb{U}[.0495, .0505]$, $\zeta''_i\in\mathbb{U}[.1995, .2005]$, $\delta=.25$, and $T=400$ in the ES; $L=3$ in the SC; $n_s=4$ in the RC; and $\lambda=.33$ and $\mu=.33$ in the CF.  We consider 41 safety levels $\tau \in \{k/40|k=0,\dots, 40\}$. For each safety level $\tau$, our simulation generates cow groups $\{N^{(t)}_l | l = 1,2,3 \ \text{and} \ t = 1, \dots, 400\}$, where $|N^{(t)}_j|=n^{(t)}_l$ allows us to compute the mean group size and the standard deviation of the group sizes. We perform similar simulations for $\sigma_x=\sigma_y=.05$, $\sigma_x=\sigma_y=.2$, and $\sigma_x=\sigma_y=.6$ with same initial states, initial conditions, and parameter values (other than the coupling strengths). We then compute the mean group size and standard deviation with respect to the safety size in each case.

In Fig.~\ref{fig:orbit}(b), we plot the mean group size and group-size standard deviation for each choice of coupling strengths. For each choice, we observe that the mean group size is about $4$ for $\tau=0$ and $\tau\in[.8, 1]$, it increases for $\tau\in(0, .58]$, and it decreases for $\tau\in(.6, .8)$. We observe [see Eq.~(\ref{eqn:risk2})] that the herd can maintain a small risk even when splitting into small groups for small safety levels. When the safety level increases, the mean group size increases to ensure that the cost of the RC is sufficiently small. However, beyond some value of the mean group size, the SC starts to dominate the CF. The mean group size thus starts to decreases for larger safety levels. In this example, the value of the safety level at which this trade-off balances is about $\tau = 0.6$. We also observe that increasing the coupling strength increases the mean group size. For large safety levels, cows can form large groups with similar dynamics without the herd incurring a significant cost.

\subsection{Cost function}

We now examine how the cost changes with respect to four parameters: the coupling strengths $\sigma_x$ and $\sigma_y$, the safe size $n_s$, and the safety level $\tau$. We perform three numerical experiments: one to examine the effect of coupling strengths; another to examine the effects of the safe size and safety levels, and another to compare the effect of safety level for zero and nonzero coupling strengths.  

\begin{figure}[htp]
    \centering
    \vspace{20pt}
    \includegraphics[width=.5\textwidth]{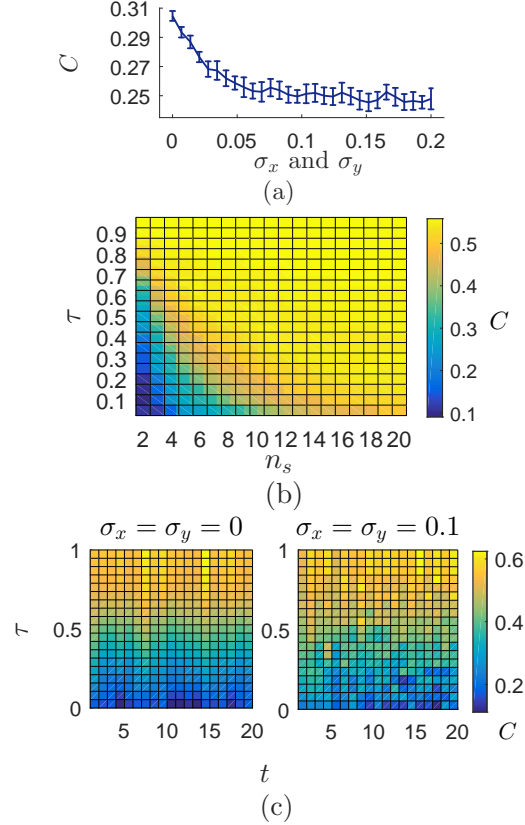}     
    \caption{Influence of parameter values on the cost $C$ of a herd of $n = 15$ cows. (a) Cost with equal coupling strengths $\sigma_x$ and $\sigma_y$, which we compute by averaging over five realizations of simulations with initial conditions from Eq.~(\ref{eqn:ini_states_ex1}) and parameters $\xi'_i\in\mathbb{U}[.0995, .1005]$, $\xi''_i\in\mathbb{U}[.1495, .1505]$, $\zeta'_i\in\mathbb{U}[.0495, .0505]$, $\zeta''_i\in\mathbb{U}[.1995, .2005]$, $\delta=.25$, $T=20$, $L=3$, $n_s=4$, $\tau=.2$, $\lambda=.33$, and $\mu=.33$. The error bars indicate the standard deviations over the five realizations. (b) Cost of the herd versus the safe size $n_s$ and safety level $\tau$ for $\sigma_x=\sigma_y=.1$ and the same values of $\xi'$, $\xi''$, $\zeta'$, $\zeta''$, $\delta$, $\lambda$, and $\mu$ as in panel (a). 
    (c) Temporal variation of the cost for different safety levels for (left) uncoupled cows and (right) coupled cows. The parameters $\xi'$, $\xi''$, $\zeta'$, $\zeta''$, $\delta$, $n_s$, $\lambda$, and $\mu$ are the same as in panel (a). For a given $\tau$, the cost for $n_s=4$ in panel (b) is the mean of the cost over all of the time steps at that value of $\tau$ in the right plot of panel (c).}
\label{fig:cost_para}
\end{figure}

We have already seen in Sec.~\ref{sec:dynamics} that different coupling strengths yield different dynamics for state switches in cows. From a biological perspective, a larger coupling strength corresponds to stronger interactions between cows, and we wish to explore how different interaction strengths affect the cost of synchronizing behavior. In our simulations, we average the cost over five realizations of parameter values of herds of $n = 15$ cows. Specifically, we generate five sets of initial states using $\theta^{(0)}_i\in\mathbb{U}\left\{\smc{E}, \smc{R}, \smc{S}\right\}$ for $i \in \{1, \dots, 15\}$ and then use Eq.~\eqref{eqn:ini_states_ex1} to generate five initial conditions. We then generate five sets of values for the parameters $\xi'_i\in\mathbb{U}[.0995, .1005]$, $\xi''_i\in\mathbb{U}[.1495, .1505]$, $\zeta'_i\in\mathbb{U}[.0495, .0505]$, and $\zeta''_i\in\mathbb{U}[.1995, .2005]$. We set $\delta=.25$, $n = 15$, and $T=20$ in the ES and consider the coupling strengths $\sigma_x=\sigma_y=k/150$, where $k \in \{1, \dots, 30\}$.  We also set $L=3$ in the SC; $n_s=4$ and $\tau=.2$ in the RC; and $\lambda=.33$ and $\mu=.33$ in the CF. We then run the dynamics for each initial condition and compute five cost values for each choice of coupling strengths $\sigma_x=\sigma_y=k/150$. In Fig.~\ref{fig:cost_para}(a), we plot the standard deviations of the costs, and we observe that it decreases with $\sigma_x$ (and hence with $\sigma_y$) until appearing to saturate once the coupling strength reaches a value of about $0.065$. 

Our model assesses the effect of risk in the cost using the RC. The risk levels in the RC depend on both the safety level $\tau$ and the safe size $n_s$ (see Eq.~\eqref{eqn:risk2}). In risky environments, we expect the values $\tau$ and $n_s$ to be larger than in safe environments. To assess the influence of these parameters on the CF, we perform simulations with the initial states $\theta^{(0)}_i\in\mathbb{U}\left\{\smc{E}, \smc{R}, \smc{S}\right\}$ for $i \in \{1, \dots, 15\}$ and determine the other initial conditions from Eq.~\eqref{eqn:ini_states_ex1}. 
We use $\xi'_i\in\mathbb{U}[.0995, .1005]$, $\xi''_i\in\mathbb{U}[.1495, .1505]$, $\zeta'_i\in\mathbb{U}[.0495, .0505]$,$\zeta''_i\in\mathbb{U}[.1995, .2005]$, $\delta=.25$, $\sigma_x=0.1$, $\sigma_y=0.1$, $n=20$, and $T=20$. Additionally, we let $L=3$ in the SC; $n_s=k_1$ (with $k_1 \in \{2, \dots, 20\}$) and $\tau=0.05k_2$ (with $k_2 \in \{1, \dots, 19\}$) in the RC; and $\lambda=.33$ and $\mu=.33$ in the CF. For each $(k_1, k_2) \in \{2, \dots, 20\}\times \{1, \dots, 19\}$, we perform one realization (so we only consider value of each parameter determined from a probability distribution). For each $\tau$ and $n_s$, we run the simulation for $T=20$ times steps and consider the cost value at each time. In Fig.~\ref{fig:cost_para}(b), we show the mean cost as a function of safe size $n_s$ and safety level $\tau$. We observe that the cost is low for small values of $n_s$ and $\tau$ and that it increases with increasing parameter values until it appears to saturate.

In our two experiments above, we examined how the CF depends on coupling strength and safety level. We now examine the temporal variation of the CF versus the safety level for both uncoupled cows and coupled cows. We generate one set of initial states using $\theta^{(0)}_i\in\mathbb{U}\left\{\smc{E}, \smc{R}, \smc{S}\right\}$ for $i \in \{1, \dots, 15\}$ and one set of initial values using Eq.~\eqref{eqn:ini_states_ex1}. We then choose the parameters in the ES, SC, and CF as in our simulations to examine the influence of coupling strengths on the CF. We set $\sigma_x=0$ and $\sigma_y=0$ for the uncoupled cows and $\sigma_x=0.1$ and $\sigma_y=0.1$ to examine a situation with coupled cows. In the RC, we let $n_s=4$ and consider a safety level of $\tau = k/25$, where $k=0, \dots, 25$. In Fig.~\ref{fig:cost_para}(c), we show the cost as a function of time and safety level for both uncoupled and coupled cows. We observe for uncoupled cows that the cost is larger for a larger safety level. Importantly, however, this need not be the case for coupled cows.

\section{Biologically-motivated examples}\label{sec:examples}

We examine the CF \eqref{cost2} using two biological examples: (1) a herd that splits into up to three groups and (2) a herd with males and females that splits into two groups.

\subsection{Example 1}\label{sec:example01}

In this example, we illustrate a scenario of a herd splitting into up to three groups. It also helps convey the effect of choosing parameter values in Eq.~(\ref{cost2}) and the relationship between groupings and their associated costs.

We consider a herd of $n=12$ cows that we allow to split into a maximum of $L=3$ groups during $T=30$ time steps. We first simulate hunger and lying desire, then compute the CF, and finally optimize the CF to determine the lowest-cost grouping at each time. We consider a single realization of the model (i.e., one example herd) and use it to illustrate the general notion of trade-offs in the CF. 

In the ES, we set the initial states of cows to be $\theta^{(0)}_i\in\mathbb{U}\{\smc{E}, \smc{R}, \smc{S}\}$ for $i \in \{1, \dots, n\}$, and we recall that $\mathbb{U}$ denotes a uniform probability distribution over the set in its argument. We set the initial conditions $x^{(0)}_i$ and $y^{(0)}_i$ according to Eq.~\eqref{eqn:ini_states_ex1}. We also make the following parameter choices for the ES: $\xi'_i,\in\mathbb{U}[.0995, .1005]$, $\xi''_i\in\mathbb{U}[.0495, .0505]$, $\zeta'_i\in\mathbb{U}[.1245, .1255]$, $\zeta''_i\in\mathbb{U}[.0745, .0755]$, $\delta=.25$, $\sigma_x=.1$, and $\sigma_y=.1$. We set the parameters in the CF and RC to be $n_s=4$, $\tau=.2$, $\lambda=.2$, and $\mu=.2$.

\begin{table}[htp]
\caption{Possible group sizes for a herd of 12 cows that splits into a maximum of 3 groups.}
\centering
\begin{ruledtabular}
\begin{tabular}{p{22pt}| c c c c c c c c c c c c c c c c c c c}
Index & 1 & 2 & 3 & 4 & 5 & 6 & 7 & 8 & 9 & 10 & 11 & 12 & 13 & 14 & 15 & 16 & 17 & 18 & 19\\ \hline
$n^{(t)}_1$ &12 & 11 & 10 & 9 & 8 & 7 & 6 & 10 & 9 & 8 & 7 & 6 & 8 & 7 & 6 & 5 & 6 & 5 &4\\
$n^{(t)}_2$ & 0 & 1 & 2 & 3 & 4 & 5 & 6 & 1 & 2 & 3 & 4 & 5 & 2 & 3 & 4 & 5 & 3 & 4 & 4\\
$n^{(t)}_3$ & 0 & 0 & 0 & 0 & 0 & 0 & 0 & 1 & 1 & 1 & 1 & 1 & 2 & 2 & 2 & 2 & 3 & 3 & 4\\
\end{tabular} \label{tbl:groups}
\end{ruledtabular}
\end{table}

\begin{figure*}[hpt]
        	\centering
        	\includegraphics[width=.8\textwidth]{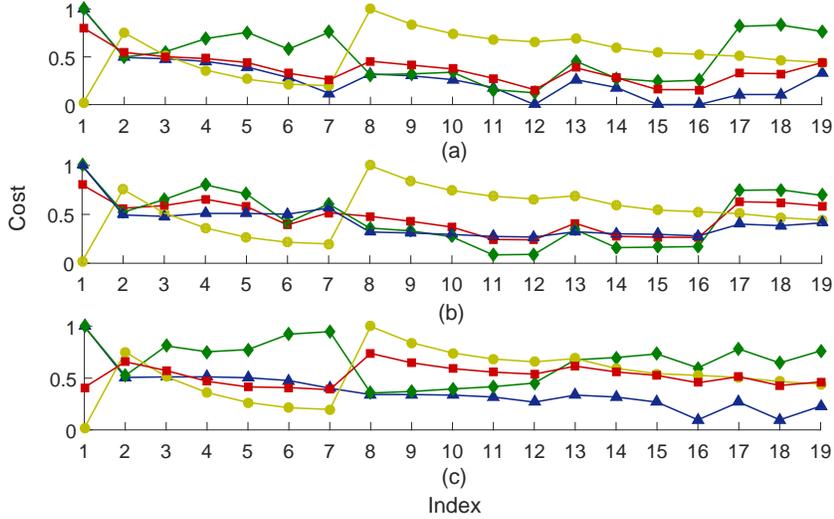}
        	\caption{Cost for different combinations of group sizes for the dynamics of a group of $n = 12$ cows for $T=30$ time steps. The parameter values are $\xi'_i\in\mathbb{U}[.0995, .1005]$, $\xi''_i\in\mathbb{U}[.0495, .0505]$, $\zeta'_i\in\mathbb{U}[.1245, .1255]$, $\zeta''_i\in\mathbb{U}[.0745, .0755]$, $\delta=.25$, $\sigma_x=.1$, and $\sigma_y=.1$ in the ES; $L=3$ in the SC; $n_s=4$ and $\tau=.2$ in the RC; and $(\lambda, \mu) \in \{(.6, .2), (.2, .6), (.2, .2)\}$ in the CF. We show the total cost (red squares) at time $t=20$ and its components --- hunger (blue triangles), lying desire (green diamonds), and risk (yellow disks) --- versus the index that represents the different combinations of group sizes (see Table \ref {tbl:groups}). The CF parameter values are (a) $\lambda=.6$ and $\mu=.2$, (b) $\lambda=.2$ and $\mu=.6$, and (c) $\lambda=.2$ and $\mu=.2$.}
	\label{fig:ex01_cost_components}
\end{figure*}

A herd of 12 cows can split into a maximum of 3 groups in 19 different combinations of group sizes (see Table \ref{tbl:groups}). We assign an index for each combination to simplify the labeling in our subsequent figures. We also run the ES together with the CF for another two instances of the CF parameters: $\lambda=.6$, $\mu=.2$ and $\lambda=.2$, $\mu=.6$.  We show our results at time $t=20$ for all three examples in Fig.~\ref{fig:ex01_cost_components}. In the figure, the highest risk occurs for $n^{(20)}_{1,2,3}=10,1,1$, in which two individual cows have separated from a herd. The second-highest risk occurs when $n^{(20)}_{1,2,3}=0,1,11$, in which one cow has separated from a herd. The lowest risk occurs when the entire herd stays together (index 1) or when it splits into equal groups (index 7), where we note that the group size of 6 is larger than the safety size $n_s = 4$. One can consider equally-weighted cost components in the convex combination that constitutes the CF or change the importance of components by increasing the weight of hunger [see Fig.~\ref{fig:ex01_cost_components}(a)], lying desire [see Fig.~\ref{fig:ex01_cost_components}(b)], or risk [see Fig.~\ref{fig:ex01_cost_components}(c)].

\begin{figure*}[hpt]
        	\centering
        	\includegraphics[width=1\textwidth]{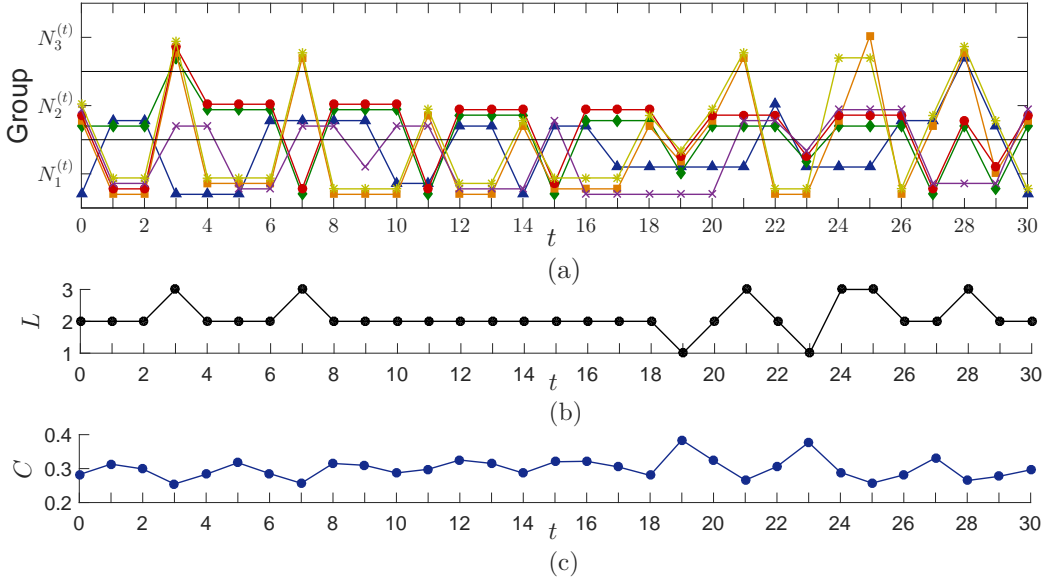}
        	\caption{Group changes and related costs as a function of time for a group of $n = 12$ cows for $T=30$ time steps. We use the parameter values $\xi'_i\in\mathbb{U}[.0995, .1005]$, $\xi''_i\in\mathbb{U}[.0495, .0505]$, $\zeta'_i\in\mathbb{U}[.1245, .1255]$, $\zeta''_i\in\mathbb{U}[.0745, .0755]$, $\delta=.25$, $\sigma_x=.1$, and $\sigma_y=.1$ in the ES; $L=3$ in the SC; $n_s=4$ and $\tau=.2$ in the RC; and $\lambda=.33$ and $\mu=.33$ in the CF.  (a) Group assignments $N^{(t)}_{1,2,3}$ of six cows (red disk, orange square, yellow asterisk, blue triangle, green diamond, and purple cross) among three groups. The (b) number of groups in which the herd splits is determined by (c) the total cost.}
	\label{fig:ex01_group_structure}
\end{figure*}

Let's now examine the temporal grouping in the scenario with parameter values $\lambda=.33$ and $\mu=.33$. In Fig.~\ref{fig:ex01_group_structure}(a), representing 6 arbitrary cows out of 12 in total, we see that cows freely switch their groups to achieve the optimum value of the CF \eqref{cost2}. The cow that we represent with the purple crosses switches between two groups during the entire simulation, whereas the other five cows switch between all three groups. In Fig.~\ref{fig:ex01_group_structure}(b), we show the total number of groups in the herd, which consists of a single group at times $t=19$ and $t=23$ and consists of three groups at times $t= 3$, $t=7$, $t=21$, $t=24$, $t=25$, and $t=28$. In Fig.\ref{fig:ex01_group_structure}(c), we show the total cost and thereby reveal that it can be more costly for the herd to stay together as a single group than to split up (at times $t = 19$ and $t = 23$). We also note the low costs for times $t = 3$, $t=7$, $t=21$, $t=25$, and $t=28$, when the herd consists of three groups. Note that we have illustrated trade-offs in the CF specifically for the initial condition and parameter values in our example, and we expect to see qualitatively different trade-offs for different initial conditions and parameter values. (Additionally, the ``high'' and ``low'' costs are not that different from each other.) However, the notion of such trade-offs is a rather general one.

\subsection{Example 2}\label{sec:example02}

We now examine mixed-sexual grouping dynamics in a herd that consists of two distinct categories of adult cows: males and females. This type of grouping is known to occur in some animal groups (e.g., red deer \cite{conradt2000activity}), so we study the same phenomenon in our model of cow herds. Adult male cows require more energy and rest than female cows, as the former tend to have larger body weights.\cite{frisch1977food, illius1987allometry} We therefore assume that the males' rates of change of hunger and lying desire are larger than those of females. Mathematically, we implement this assumption by using larger values of the parameters $\xi'_i, \xi''_i, \zeta'_i$, and $\zeta''_i$ of cows in the male group than for those in the female group.

We consider a herd of 10 cows that consists of two groups. The first group has five cows (where $i \in \{1, \dots, 5\}$ indexes the cow) with large body weights, and the second group has the remaining five cows ($i \in \{6, \dots, 10\}$), which have small body weights. As in Sec.~\ref{sec:example01}, we simulate the hunger and lying desire of cows with the ES (see Sec.~\ref{sec:temporal_evolution}) and determine a lowest-cost grouping by optimizing the CF \eqref{cost2}. We set the initial states of cows in the first and second groups as eating and lying down, respectively. Within a group, the variables have very similar initial values. Specifically, they are the same, except that we perturb them additively with a small amount of uniform noise:
\begin{equation}\label{eqn:ex1_intial_cond}
	\left\{\def\arraystretch{1.2}%
\begin{array}{@{}c@{\quad}l@{}}
    \theta^{(0)}_i =\text{$\smc{E}$} \text{ and } \left(x^{(0)}_i, y^{(0)}_i\right)=[1, \delta+\phi_i]\,, & i\in \{1, \dots, 5\}\,,\\
    \theta^{(0)}_i =\text{$\smc{R}$} \text{ and } \left(x^{(0)}_i, y^{(0)}_i\right)=[\delta+ \phi '_i, 1]\,, & i\in \{6, \dots, 10\}\,,\\
\end{array}\right.
\end{equation}
where $\phi_i, \phi'_i \in10^{-3} \mathbb{U}[0, 1]$ and $\delta=.25$. We choose uniform additive noise because it is the simplest type of noise to consider. We set $\sigma_x=.2$ and $\sigma_y=.2$ in the ES, and we determine the other parameters so that the first group consists of cows with large body mass and the second group consists of cows with small body mass:
\begin{equation}\label{eqn:parameters}
	\left\{\def\arraystretch{1.4}%
\begin{array}{@{}c@{\quad}l@{}}
    \left\{\def\arraystretch{1.4}%
    \begin{array}{@{}c@{\quad}l@{}}
         \xi'_i, \zeta'_i\in\mathbb{U}[.2495, .2505]\,,\\
         \xi''_i\in\mathbb{U}[.2995, .3005]\,,\\
         \zeta''_i\in\mathbb{U}[.3995, .4005]\,,\\	
    \end{array}\right. &  i \in \{1, \dots, 5\}\,,\\
    \left\{\def\arraystretch{1.4}% 
    \begin{array}{@{}c@{\quad}l@{}}
         \xi'_i\in\mathbb{U}[.0995, .1005]\,,\\         
         \xi''_i\in\mathbb{U}[.0495, .0505]\,,\\
         \zeta'_i\in\mathbb{U}[.1245, .1255]\,,\\
         \zeta''_i\in\mathbb{U}[.0745, .0755]\,,\\
    \end{array}\right. &  i\in \{6, \dots, 10\}\,.\\
\end{array}\right.
\end{equation}

\begin{figure*}[hpt]
        	\centering
        	\includegraphics[width=.95\textwidth]{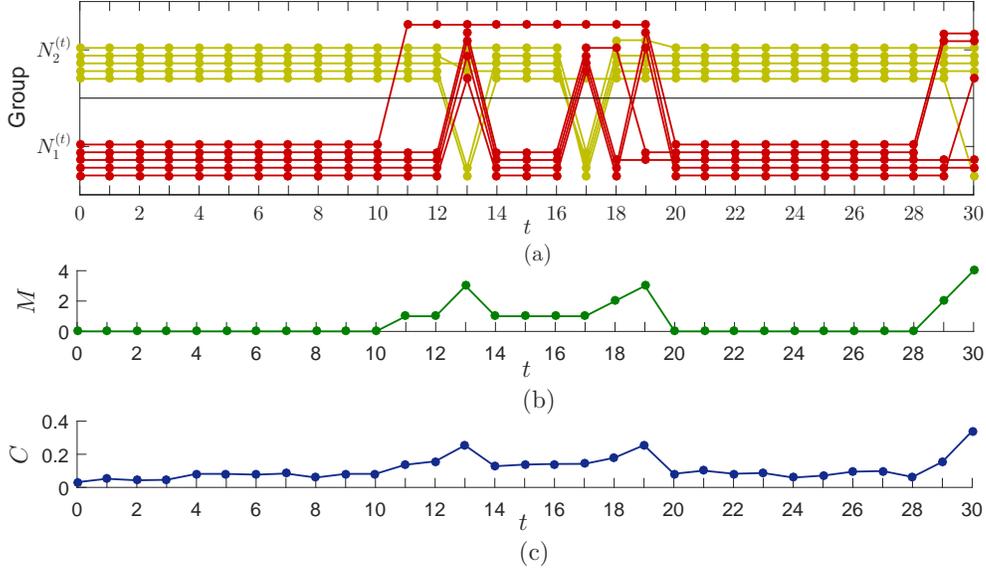}
        	\caption{Dynamics and cost function over time of a 10-cow herd of adult males and females that splits into two groups. We use the parameter values $\xi'_i, \zeta'_i\in\mathbb{U}[.2495, .2505]$, $\xi''_i\in\mathbb{U}[.2995, .3005]$, and $\zeta''_i\in\mathbb{U}[.3995, .4005]$ for the first 5 cows to create ``male'' cows; and we use the parameter values $\xi'_i\in\mathbb{U}[.0995, .1005]$, $\xi''_i\in\mathbb{U}[.0495, .0505]$, $\zeta'_i\in\mathbb{U}[.1245, .1255]$, and $\zeta''_i\in\mathbb{U}[.0745, .0755]$ for the other 5 cows to create ``female'' cows. The other parameter values are $\delta=.25$, $\sigma_x=.2$, and $\sigma_y=.2$ in the ES; $L=2$ in the SC; $n_s=3$, and $\tau=.2$ in the RC; and $\lambda=.33$ and $\mu=.33$ in the CF.  (a) Cow groups as a function of time. We color the first 5 cows (the ``male'' group) in red and the other 5 cows (the ``female'' group) in yellow. (b) Number of cows that are not in their proper group as a function of time. (c) Cost of the groups over time.}
	\label{fig:ex02_group_structures}
\end{figure*}

We set the parameters in the CF and RC to be $n_s=3$, $\tau=.2$, $\lambda=.33$, and $\mu=.33$. We run the ES for $T=30$ time steps, and we consider the value of the CF at each step. As in our example in Sec.~\ref{sec:example01}, we use only one realization, and we note that the noise in Eq.~(\ref{eqn:ex1_intial_cond}) has a small magnitude. During time steps 0--10 and 20--28, we see in Fig.~\ref{fig:ex02_group_structures}(a) that all of the cows are in groups with the other cows of their own sex (i.e., with others of similar sizes, hunger, and desire to lie down). However, during time steps 11--19 and 29--30, some cows are not in their ``proper'' group, and the cost becomes high [see Fig.~\ref{fig:ex02_group_structures}(c)], although the CF minimizes the cost to achieve a lowest-cost grouping. We show the number of mismatched cows in the groups in Fig.~\ref{fig:ex02_group_structures}(b). We observe that the cost is large when cows are in mismatched groups, but it is low when cows are in their proper (i.e., homogeneous-sex) groups.

\section{Conclusions and Discussion}\label{sec:conc_discu}

We developed a framework for modeling the lowest-cost splitting of a herd of cows by optimizing a cost function (CF) that quantifies their hunger, desire to lie down, and predation risk. Lying in groups offers protection from predators,\cite{bode2010perceived, ruckstuhl1998foraging, estevez2007group, mendl2001living}, but synchronization can also be costly to individuals, as some portion of a herd has to change behavior to eat or lie down at a communal time rather than at an optimally beneficial time. \cite{ruckstuhl1999synchronise, conradt2000activity, dostalkova2010go} In this paper, we examined situations in which cow herds split into groups such that cows' hunger and lying desire are relatively homogeneous within a group, while ensuring that further splitting does not result in overly small groups, which would be more vulnerable to predation.

We employed the evolution scheme (ES) from Sun et al.~\cite{sun2011mathematical} and input cows' time-dependent interactions in terms of a adjacency matrix $A^{(t)}$ that encodes a lowest-cost grouping obtained by optimizing the CF. The adjacency matrix provides an interface between the CF and ES, and our framework can be used with arbitrarily intricate CFs, ESs, and interaction patterns. In Ref.~\onlinecite{sun2011mathematical}, the network architecture $A^{(t)}$, which indicates which cows are coupled to each other at each time $t$, was imposed as part of the model. In the present paper, however, we took a different approach: we determined $A^{(t)}$ based on an optimum grouping at the previous time step (after imposing a group structure at $t = 0$ as part of the initial conditions). Because hunger and lying desire are two separate motivations of a cow, we optimized the CF independently for each one in each time step to obtain two different groupings, and we then used the grouping with the lower total cost among the two possibilities. For convenience, we imposed a maximum number of groups into which a herd can split, as it reduces the computational complexity of our approach. We assessed the cost contributions from hunger and lying desire using the standard deviation of the associated individual preferences in each group (although one can replace the standard deviation by any measure of dispersion).

In Sec.~\ref{sec:analysis}, we first examined how cow dynamics are affected by coupling strengths, and we then examined the CF for different parameter values. We simulated hunger and lying desire of cows for four sets of coupling strengths and observed different dynamics [see Fig.~\ref{fig:orbit}(a)] in the four situations. Setting the coupling strengths to $0$ implies that each cow behaves independently [see Eq.~\eqref{eqn:diff_equation}], so each cow oscillates as an independent oscillator. In contrast, for positive coupling strengths, cows interact with each other, and we represent a cow herd as a set of coupled oscillators. To examine the different dynamics from different coupling strengths in a biologically-motivated context, we computed the mean group size versus the safety level for different coupling strengths. We observed in Fig.~\ref{fig:orbit}(b) that large coupling strengths permit large groups that consist of cows with similar needs. We also observed that the mean group size of cows first increases when the safety level increases but then decreases after some value of the safety level [see Fig.~\ref{fig:orbit}(b)]. Recall that group sizes in a herd also increase with the safety level [see Eq.~\eqref{fig:risk_comp}]. In sufficiently large groups, the synchronization cost starts to dominate the CF for sufficiently large safety levels and minimizing the CF starts to encourage smaller groups to minimize the cost. Thereafter, the mean group size decreases with safety level.

We then studied the influence of coupling strengths $\sigma_x$ and $\sigma_y$, safe size $n_s$, and safety level $\tau$ on the CF. We observed [see Fig.~\ref{fig:cost_para}(a)] that the the total cost decreases with increased coupling strength before saturating. In Fig.~\ref{fig:cost_para}(b), we illustrated that setting the safe size and safety level to low values entails a low cost. Such low parameter values allow cows to gather into small groups of similar cows without incurring a significant risk to a herd. When the cows are uncoupled, the cost increases monotonically with increasing safety level, but the cost varies non-monotonically with increasing safety levels for coupled cows [see Fig.~\ref{fig:cost_para}(c)].

In a biologically-motivated example, we examined group fission and the dynamics of cows switching between groups. In that example, we set the initial states of cows arbitrarily, but one can also choose initial states to examine specific scenarios. To consider a relatively homogeneous herd, we used similar parameter values for different individuals, and we observed the dynamics that result from small differences in these parameter values. We considered a single realization of the model, as the other initial conditions and parameter values would yield different specific trade-offs but illustrate the same essential idea. Our primary hypothesis, that synchronization can be costly, is illustrated by Figs.~\ref{fig:ex01_group_structure}(b,c). Specifically, synchronization is very costly when the groups are large and heterogeneous. One could explore trade-offs further by considering risk and synchronization costs with different rates of increase with group size.

One can customize the ES by changing the parameters for the rates of increase in hunger or desire to lie down. This versatility allowed us to model a scenario of mixed-sexual grouping in a herd. Adult male cows generally possess larger body masses and require more energy and lying time than adult female cows. We implemented this asymmetry among individuals by imposing larger values of the salient parameters for males than for females. At times, the heterogeneity in motivations for eating and lying down caused the optimal groups to consist of cow groups other than the single-sex groups [see Fig.~\ref{fig:ex02_group_structures}(c)], but usually optimization of the CF yielded single-sex groups. Single-sex grouping occurs commonly in ungulates (e.g., cows, deer, and sheep) and are especially pronounced in species with large body-size differences between males and females.\cite{ruckstuhl1999synchronise, conradt2000activity, ruckstuhl2002sexual, conradt2009group} In our exploration of sex grouping, we added uniform noise to the initial conditions and parameters, as it is the simplest type of noise to consider. 

One can adjust the CF so that it can be used for herding situations in different environments. A safe environment allows small groups in a herd, in contrast to an unsafe environment, which requires large groups to defend themselves against attacks. Our CF imitates a safe environment if the safe size $n_s$ is large and the safety level is small. One can control the influence of the cost components (hunger, lying desire, and predation risk) on the CF by tuning parameters, and our approach thereby makes it possible to explore different grouping scenarios, such as analyzing the influence of one or more cost components over the others for group splitting. Our overall approach is also very flexible, and one can generalize our CF, the ES, and the interactions among animals (through a time-dependent adjacency matrix) to examiner a wide variety of scenarios.

In our paper, we determined group size and splitting by optimizing a CF at each time step. However, because optimally-sized groups are not necessarily stable, it may be necessary in future work to introduce a learning process in which one keeps track of optimal group sizes during past time steps. In the present paper, we imposed a maximum number $L$ of groups into which a herd can split. In our examples, the value of $L$ was either obvious, as in the sex-grouping example, where we used $L = 2$ (males and females), or hypothetical, as in our example with $L=3$. However, instead of imposing a maximum number of groups in advance, it is also desirable to examine situations in which the number of groups is an unconstrained output to better reveal an optimal number of groups in herd splitting.

In summary, we developed a versatile model of lowest-cost splitting of a herd of animals that can admits numerous generalizations in a straightforward way. We illustrated our model by exploring several plausible scenarios, and we believe that our approach has the potential to shed considerable insight on grouping behavior in animals in a wide variety of situations.

\section*{Acknowledgements}

We thank Jie Sun for valuable comments on this work. EMB and KG were supported by the National Science Foundation (DMS-0404778), and EMB was also supported by the Office of Naval Research (N00014-15-1-2093) and the Army Research Office (N68164-EG and W911NF-12-1- 0276).

\appendix

\section{Derivation of the discrete dynamics on the Poincar\'e section}\label{sec:appendix}

We solve the differential equations in Eq.~\eqref{eqn:boundaries} using the boundary conditions in Eq.~\eqref{eqn:state_change}. For convenience, we substitute Eq.~\eqref{eqn:subs_diff} into these differential equations and expand as follows:\\
when $\theta^{(t)}_i=\smc{E}$,
\begin{equation}\label{eqn:d1}
	\begin{split}
\dot x^{(t+1)}_i=\eta''_i x^{(t)}_i\,, \\
\dot y^{(t+1)}_i=\gamma'_i y^{(t)}_i\,;
	\end{split}
\end{equation}
when $\theta^{(t)}_i=\smc{R}$,
\begin{equation}\label{eqn:d2}
	\begin{split}
\dot x^{(t+1)}_i=\eta'_i x^{(t)}_i\,, \\
\dot y^{(t+1)}_i=\gamma''_i y^{(t)}_i\,;
	\end{split}
\end{equation}
when $\theta^{(t)}_i=\smc{E}$,
\begin{equation}\label{eqn:d3}
	\begin{split}
\dot x^{(t+1)}_i=\eta'_i x^{(t)}_i\,, \\
\dot y^{(t+1)}_i=\gamma'_i y^{(t)}_i\,.
	\end{split}
\end{equation}

We then solve the differential equations in Eqs.~\eqref{eqn:d1}--\eqref{eqn:d3} on the boundaries $\partial \mathcal{E}$, $\partial \mathcal{R}$, $\partial \mathcal{S}_x$, and $\partial \mathcal{S}_y$ given by Eq.~\eqref{eqn:boundaries} as follows:\\\\\\
when $\theta^{(t)}_i=\smc{E}$ (i.e., on $\partial \mathcal{E}$ of the Poincar\'e section),
\begin{equation}\label{eqn:ds1}
	\left\{\def\arraystretch{1.2}%
\begin{array}{@{}l@{\quad}l@{}}
    t_{\mathcal{E}\mathcal{R}} =\frac{1}{\gamma'_i} \log\Big(\frac{1}{y^{(t)}_i}\Big)\,, \text{ so}\\ 
\hspace{1.5cm}\left(x^{(t+1)}_i, y^{(t+1)}_i, \theta^{(t+1)}_i\right)=\left(\Big(y^{(t)}_i\Big)^{\frac{\eta''_i}{\gamma'_i}}, 1, \mathcal{R}\right)\,;\\
    t_{\mathcal{E}\mathcal{S}_y} =\frac{1}{\eta''_i} \log\Big(\frac{1}{\delta}\Big)\,, \text{ so}, \\
\hspace{1.5cm}\left(x^{(t+1)}_i, y^{(t+1)}_i, \theta^{(t+1)}_i\right)=\left(\delta, \delta^{-\frac{\gamma'_i}{\eta''_i}}y^{(t)}_i, \mathcal{S}_y\right);\\
\end{array}\right.
\end{equation}\\\\
when $\theta^{(t)}_i=\smc{R}$ (i.e., on $\partial \mathcal{R}$ of the Poincar\'e section),
\begin{equation}\label{eqn:ds2}
	\left\{\def\arraystretch{1.2}%
\begin{array}{@{}l@{\quad}l@{}}
    t_{\mathcal{R}\mathcal{E}} =\frac{1}{\eta'_i} \log\Big(\frac{1}{x^{(t)}_i}\Big)\,, \text{ so}\\
\hspace{1.5cm}\left(x^{(t+1)}_i, y^{(t+1)}_i, \theta^{(t+1)}_i\right)=\left(1, \Big(x^{(t)}_i\Big)^{\frac{\gamma''_i}{\eta'_i}}, \mathcal{E}\right)\,;\\
    t_{\mathcal{R}\mathcal{S}_x} =\frac{1}{\gamma''_i} \log\Big(\frac{1}{\delta}\Big)\,, \text{ so}\\
\hspace{1.5cm}\left(x^{(t+1)}_i, y^{(t+1)}_i, \theta^{(t+1)}_i\right)=\left(\delta^{-\frac{\eta'_i}{\gamma''_i}}x^{(t)}_i, \delta, \mathcal{S}_x\right)\,;\\
\end{array}\right.
\end{equation}\\\\
when $\theta^{(t)}_i=\smc{S}_y$ (i.e., on $\partial \mathcal{S}_y$ of the Poincar\'e section),
\begin{equation}\label{eqn:ds3}
	\left\{\def\arraystretch{1.2}%
\begin{array}{@{}l@{\quad}l@{}}
    t_{\mathcal{S}_y\mathcal{E}} =\frac{1}{\eta'_i} \log\Big(\frac{1}{\delta}\Big)\,, \text{ so}\\
\hspace{1.5cm}\left(x^{(t+1)}_i, y^{(t+1)}_i, \theta^{(t+1)}_i\right)=\left(1, \delta^{-\frac{\gamma'_i}{\eta'_i}}y^{(t)}_i, \mathcal{E}\right)\,;\\
    t_{\mathcal{S}_y\mathcal{R}} =\frac{1}{\gamma'_i} \log\Big(\frac{1}{y^{(t)}_i}\Big)\,, \text{ so}\\
\hspace{1.5cm}\left(x^{(t+1)}_i, y^{(t+1)}_i, \theta^{(t+1)}_i\right)=\left(\Big(y^{(t)}_i\Big)^{-\frac{\eta'_i}{\gamma'_i}}\delta, 1, \mathcal{R}\right)\,;\\
\end{array}\right.
\end{equation}\\\\
when $\theta^{(t)}_i=\smc{S}_x$ (i.e., on $\partial \mathcal{S}_x$ of the Poincar\'e section),
\begin{equation}\label{eqn:ds4}
	\left\{\def\arraystretch{1.2}%
\begin{array}{@{}l@{\quad}l@{}}
    t_{\mathcal{S}_x\mathcal{E}} =\frac{1}{\eta'_i} \log\Big(\frac{1}{x^{(t)}_i}\Big)\,, \text{ so}\\
\hspace{1.5cm}\left(x^{(t+1)}_i, y^{(t+1)}_i, \theta^{(t+1)}_i\right)=\left(1, \Big(x^{(t)}_i\Big)^{-\frac{\gamma'_i}{\eta'_i}}\delta, \mathcal{E}\right)\,;\\
    t_{\mathcal{S}_x\mathcal{R}} =\frac{1}{\gamma'_i} \log\Big(\frac{1}{\delta}\Big)\,, \text{ so}\\
\hspace{1.5cm}\left(x^{(t+1)}_i, y^{(t+1)}_i, \theta^{(t+1)}_i\right)=\left(\delta^{-\frac{\eta'_i}{\gamma'_i}}x^{(t)}_i, 1, \mathcal{R}\right)\,.\\
\end{array}\right.
\end{equation}

%%%%%
%\bibliography{ref}

\begin{thebibliography}{53}%
\makeatletter
\providecommand \@ifxundefined [1]{%
 \@ifx{#1\undefined}
}%
\providecommand \@ifnum [1]{%
 \ifnum #1\expandafter \@firstoftwo
 \else \expandafter \@secondoftwo
 \fi
}%
\providecommand \@ifx [1]{%
 \ifx #1\expandafter \@firstoftwo
 \else \expandafter \@secondoftwo
 \fi
}%
\providecommand \natexlab [1]{#1}%
\providecommand \enquote  [1]{``#1''}%
\providecommand \bibnamefont  [1]{#1}%
\providecommand \bibfnamefont [1]{#1}%
\providecommand \citenamefont [1]{#1}%
\providecommand \href@noop [0]{\@secondoftwo}%
\providecommand \href [0]{\begingroup \@sanitize@url \@href}%
\providecommand \@href[1]{\@@startlink{#1}\@@href}%
\providecommand \@@href[1]{\endgroup#1\@@endlink}%
\providecommand \@sanitize@url [0]{\catcode `\\12\catcode `\$12\catcode
  `\&12\catcode `\#12\catcode `\^12\catcode `\_12\catcode `\%12\relax}%
\providecommand \@@startlink[1]{}%
\providecommand \@@endlink[0]{}%
\providecommand \url  [0]{\begingroup\@sanitize@url \@url }%
\providecommand \@url [1]{\endgroup\@href {#1}{\urlprefix }}%
\providecommand \urlprefix  [0]{URL }%
\providecommand \Eprint [0]{\href }%
\providecommand \doibase [0]{http://dx.doi.org/}%
\providecommand \selectlanguage [0]{\@gobble}%
\providecommand \bibinfo  [0]{\@secondoftwo}%
\providecommand \bibfield  [0]{\@secondoftwo}%
\providecommand \translation [1]{[#1]}%
\providecommand \BibitemOpen [0]{}%
\providecommand \bibitemStop [0]{}%
\providecommand \bibitemNoStop [0]{.\EOS\space}%
\providecommand \EOS [0]{\spacefactor3000\relax}%
\providecommand \BibitemShut  [1]{\csname bibitem#1\endcsname}%
\let\auto@bib@innerbib\@empty
%</preamble>
\bibitem [{\citenamefont {Elgar}(1989)}]{elgar1989predator}%
  \BibitemOpen
  \bibfield  {author} {\bibinfo {author} {\bibfnamefont {M.~A.}\ \bibnamefont
  {Elgar}},\ }\bibfield  {title} {\enquote {\bibinfo {title} {Predator
  vigilance and group size in mammals and birds: {A} critical review of the
  empirical evidence},}\ }\href@noop {} {\bibfield  {journal} {\bibinfo
  {journal} {Biological Reviews}\ }\textbf {\bibinfo {volume} {64}},\ \bibinfo
  {pages} {13--33} (\bibinfo {year} {1989})}\BibitemShut {NoStop}%
\bibitem [{\citenamefont {Krause}\ and\ \citenamefont
  {Ruxton}(2002)}]{krause2002living}%
  \BibitemOpen
  \bibfield  {author} {\bibinfo {author} {\bibfnamefont {J.}~\bibnamefont
  {Krause}}\ and\ \bibinfo {author} {\bibfnamefont {G.~D.}\ \bibnamefont
  {Ruxton}},\ }\href@noop {} {\emph {\bibinfo {title} {Living in Groups}}}\
  (\bibinfo  {publisher} {Oxford University Press},\ \bibinfo {year}
  {2002})\BibitemShut {NoStop}%
\bibitem [{\citenamefont {Brown}\ and\ \citenamefont
  {Brown}(1986)}]{brown1986ectoparasitism}%
  \BibitemOpen
  \bibfield  {author} {\bibinfo {author} {\bibfnamefont {C.~R.}\ \bibnamefont
  {Brown}}\ and\ \bibinfo {author} {\bibfnamefont {M.~B.}\ \bibnamefont
  {Brown}},\ }\bibfield  {title} {\enquote {\bibinfo {title} {Ectoparasitism as
  a cost of coloniality in cliff swallows (\emph{Hirundo pyrrhonota})},}\
  }\href@noop {} {\bibfield  {journal} {\bibinfo  {journal} {Ecology}\ }\textbf
  {\bibinfo {volume} {67}},\ \bibinfo {pages} {1206--1218} (\bibinfo {year}
  {1986})}\BibitemShut {NoStop}%
\bibitem [{\citenamefont {Altizer}\ \emph {et~al.}(2003)\citenamefont
  {Altizer}, \citenamefont {Nunn}, \citenamefont {Thrall}, \citenamefont
  {Gittleman}, \citenamefont {Antonovics}, \citenamefont {Cunningham},
  \citenamefont {Cunnningham}, \citenamefont {Dobson}, \citenamefont {Ezenwa},
  \citenamefont {Jones} \emph {et~al.}}]{altizer2003social}%
  \BibitemOpen
  \bibfield  {author} {\bibinfo {author} {\bibfnamefont {S.}~\bibnamefont
  {Altizer}}, \bibinfo {author} {\bibfnamefont {C.~L.}\ \bibnamefont {Nunn}},
  \bibinfo {author} {\bibfnamefont {P.~H.}\ \bibnamefont {Thrall}}, \bibinfo
  {author} {\bibfnamefont {J.~L.}\ \bibnamefont {Gittleman}}, \bibinfo {author}
  {\bibfnamefont {J.}~\bibnamefont {Antonovics}}, \bibinfo {author}
  {\bibfnamefont {A.~A.}\ \bibnamefont {Cunningham}}, \bibinfo {author}
  {\bibfnamefont {A.~A.}\ \bibnamefont {Cunnningham}}, \bibinfo {author}
  {\bibfnamefont {A.~P.}\ \bibnamefont {Dobson}}, \bibinfo {author}
  {\bibfnamefont {V.}~\bibnamefont {Ezenwa}}, \bibinfo {author} {\bibfnamefont
  {K.~E.}\ \bibnamefont {Jones}},  \emph {et~al.},\ }\bibfield  {title}
  {\enquote {\bibinfo {title} {Social organization and parasite risk in
  mammals: {I}ntegrating theory and empirical studies},}\ }\href@noop {}
  {\bibfield  {journal} {\bibinfo  {journal} {Annual Rev. of Ecology,
  Evolution, and Systematics}\ ,\ \bibinfo {pages} {517--547}} (\bibinfo {year}
  {2003})}\BibitemShut {NoStop}%
\bibitem [{\citenamefont {Di~Bitetti}\ and\ \citenamefont
  {Janson}(2001)}]{di2001social}%
  \BibitemOpen
  \bibfield  {author} {\bibinfo {author} {\bibfnamefont {M.~S.}\ \bibnamefont
  {Di~Bitetti}}\ and\ \bibinfo {author} {\bibfnamefont {C.~H.}\ \bibnamefont
  {Janson}},\ }\bibfield  {title} {\enquote {\bibinfo {title} {Social foraging
  and the finder's share in capuchin monkeys, \emph{Cebus apella}},}\
  }\href@noop {} {\bibfield  {journal} {\bibinfo  {journal} {Animal Behaviour}\
  }\textbf {\bibinfo {volume} {62}},\ \bibinfo {pages} {47--56} (\bibinfo
  {year} {2001})}\BibitemShut {NoStop}%
\bibitem [{\citenamefont {Usherwood}\ \emph {et~al.}(2011)\citenamefont
  {Usherwood}, \citenamefont {Stavrou}, \citenamefont {Lowe}, \citenamefont
  {Roskilly},\ and\ \citenamefont {Wilson}}]{usherwood2011flying}%
  \BibitemOpen
  \bibfield  {author} {\bibinfo {author} {\bibfnamefont {J.~R.}\ \bibnamefont
  {Usherwood}}, \bibinfo {author} {\bibfnamefont {M.}~\bibnamefont {Stavrou}},
  \bibinfo {author} {\bibfnamefont {J.~C.}\ \bibnamefont {Lowe}}, \bibinfo
  {author} {\bibfnamefont {K.}~\bibnamefont {Roskilly}}, \ and\ \bibinfo
  {author} {\bibfnamefont {A.~M.}\ \bibnamefont {Wilson}},\ }\bibfield  {title}
  {\enquote {\bibinfo {title} {Flying in a flock comes at a cost in pigeons},}\
  }\href@noop {} {\bibfield  {journal} {\bibinfo  {journal} {Nature}\ }\textbf
  {\bibinfo {volume} {474}},\ \bibinfo {pages} {494--497} (\bibinfo {year}
  {2011})}\BibitemShut {NoStop}%
\bibitem [{\citenamefont {Creel}\ and\ \citenamefont
  {Winnie}(2005)}]{creel2005responses}%
  \BibitemOpen
  \bibfield  {author} {\bibinfo {author} {\bibfnamefont {S.}~\bibnamefont
  {Creel}}\ and\ \bibinfo {author} {\bibfnamefont {J.~A.}\ \bibnamefont
  {Winnie}},\ }\bibfield  {title} {\enquote {\bibinfo {title} {Responses of elk
  herd size to fine-scale spatial and temporal variation in the risk of
  predation by wolves},}\ }\href@noop {} {\bibfield  {journal} {\bibinfo
  {journal} {Animal Behaviour}\ }\textbf {\bibinfo {volume} {69}},\ \bibinfo
  {pages} {1181--1189} (\bibinfo {year} {2005})}\BibitemShut {NoStop}%
\bibitem [{\citenamefont {Hebblewhite}\ and\ \citenamefont
  {Pletscher}(2002)}]{hebblewhite2002effects}%
  \BibitemOpen
  \bibfield  {author} {\bibinfo {author} {\bibfnamefont {M.}~\bibnamefont
  {Hebblewhite}}\ and\ \bibinfo {author} {\bibfnamefont {D.~H.}\ \bibnamefont
  {Pletscher}},\ }\bibfield  {title} {\enquote {\bibinfo {title} {Effects of
  elk group size on predation by wolves},}\ }\href@noop {} {\bibfield
  {journal} {\bibinfo  {journal} {Canadian Journal of Zoology}\ }\textbf
  {\bibinfo {volume} {80}},\ \bibinfo {pages} {800--809} (\bibinfo {year}
  {2002})}\BibitemShut {NoStop}%
\bibitem [{\citenamefont {Sibly}(1983)}]{sibly1983optimal}%
  \BibitemOpen
  \bibfield  {author} {\bibinfo {author} {\bibfnamefont {R.~M.}\ \bibnamefont
  {Sibly}},\ }\bibfield  {title} {\enquote {\bibinfo {title} {Optimal group
  size is unstable},}\ }\href@noop {} {\bibfield  {journal} {\bibinfo
  {journal} {Animal Behaviour}\ }\textbf {\bibinfo {volume} {31}},\ \bibinfo
  {pages} {947--948} (\bibinfo {year} {1983})}\BibitemShut {NoStop}%
\bibitem [{\citenamefont {Sumpter}(2010)}]{sumpter2010collective}%
  \BibitemOpen
  \bibfield  {author} {\bibinfo {author} {\bibfnamefont {D.~J.~T.}\
  \bibnamefont {Sumpter}},\ }\href@noop {} {\emph {\bibinfo {title} {Collective
  Animal Behavior}}}\ (\bibinfo  {publisher} {Princeton University Press},\
  \bibinfo {year} {2010})\BibitemShut {NoStop}%
\bibitem [{\citenamefont {Ruckstuhl}(1999)}]{ruckstuhl1999synchronise}%
  \BibitemOpen
  \bibfield  {author} {\bibinfo {author} {\bibfnamefont {K.~E.}\ \bibnamefont
  {Ruckstuhl}},\ }\bibfield  {title} {\enquote {\bibinfo {title} {To
  synchronise or not to synchronise: {A} dilemma for young bighorn males?}}\
  }\href@noop {} {\bibfield  {journal} {\bibinfo  {journal} {Behaviour}\
  }\textbf {\bibinfo {volume} {136}},\ \bibinfo {pages} {805--818} (\bibinfo
  {year} {1999})}\BibitemShut {NoStop}%
\bibitem [{\citenamefont {Conradt}\ and\ \citenamefont
  {Roper}(2000)}]{conradt2000activity}%
  \BibitemOpen
  \bibfield  {author} {\bibinfo {author} {\bibfnamefont {L.}~\bibnamefont
  {Conradt}}\ and\ \bibinfo {author} {\bibfnamefont {T.~J.}\ \bibnamefont
  {Roper}},\ }\bibfield  {title} {\enquote {\bibinfo {title} {Activity
  synchrony and social cohesion: A fission--fusion model},}\ }\href@noop {}
  {\bibfield  {journal} {\bibinfo  {journal} {Proceedings of the Royal Society
  of London B: Biological Sciences}\ }\textbf {\bibinfo {volume} {267}},\
  \bibinfo {pages} {2213--2218} (\bibinfo {year} {2000})}\BibitemShut {NoStop}%
\bibitem [{\citenamefont {Dost{\'a}lkov{\'a}}\ and\ \citenamefont
  {{\v{S}}pinka}(2010)}]{dostalkova2010go}%
  \BibitemOpen
  \bibfield  {author} {\bibinfo {author} {\bibfnamefont {I.}~\bibnamefont
  {Dost{\'a}lkov{\'a}}}\ and\ \bibinfo {author} {\bibfnamefont
  {M.}~\bibnamefont {{\v{S}}pinka}},\ }\bibfield  {title} {\enquote {\bibinfo
  {title} {When to go with the crowd: {M}odelling synchronization of
  all-or-nothing activity transitions in grouped animals},}\ }\href@noop {}
  {\bibfield  {journal} {\bibinfo  {journal} {Journal of Theoretical Biology}\
  }\textbf {\bibinfo {volume} {263}},\ \bibinfo {pages} {437--448} (\bibinfo
  {year} {2010})}\BibitemShut {NoStop}%
\bibitem [{\citenamefont {Ruckstuhl}\ and\ \citenamefont
  {Neuhaus}(2002)}]{ruckstuhl2002sexual}%
  \BibitemOpen
  \bibfield  {author} {\bibinfo {author} {\bibfnamefont {K.~E.}\ \bibnamefont
  {Ruckstuhl}}\ and\ \bibinfo {author} {\bibfnamefont {P.}~\bibnamefont
  {Neuhaus}},\ }\bibfield  {title} {\enquote {\bibinfo {title} {Sexual
  segregation in ungulates: {A} comparative test of three hypotheses},}\
  }\href@noop {} {\bibfield  {journal} {\bibinfo  {journal} {Biological Reviews
  of the Cambridge Philosophical Society}\ }\textbf {\bibinfo {volume} {77}},\
  \bibinfo {pages} {77--96} (\bibinfo {year} {2002})}\BibitemShut {NoStop}%
\bibitem [{\citenamefont {Conradt}\ and\ \citenamefont
  {List}(2009)}]{conradt2009group}%
  \BibitemOpen
  \bibfield  {author} {\bibinfo {author} {\bibfnamefont {L.}~\bibnamefont
  {Conradt}}\ and\ \bibinfo {author} {\bibfnamefont {C.}~\bibnamefont {List}},\
  }\bibfield  {title} {\enquote {\bibinfo {title} {Group decisions in humans
  and animals: {A} survey},}\ }\href@noop {} {\bibfield  {journal} {\bibinfo
  {journal} {Philosophical Transactions of the Royal Society of London B:
  Biological Sciences}\ }\textbf {\bibinfo {volume} {364}},\ \bibinfo {pages}
  {719--742} (\bibinfo {year} {2009})}\BibitemShut {NoStop}%
\bibitem [{\citenamefont {Ramos}\ \emph {et~al.}(2015)\citenamefont {Ramos},
  \citenamefont {Petit}, \citenamefont {Longour}, \citenamefont {Pasquaretta},\
  and\ \citenamefont {Sueur}}]{ramos2015collective}%
  \BibitemOpen
  \bibfield  {author} {\bibinfo {author} {\bibfnamefont {A.}~\bibnamefont
  {Ramos}}, \bibinfo {author} {\bibfnamefont {O.}~\bibnamefont {Petit}},
  \bibinfo {author} {\bibfnamefont {P.}~\bibnamefont {Longour}}, \bibinfo
  {author} {\bibfnamefont {C.}~\bibnamefont {Pasquaretta}}, \ and\ \bibinfo
  {author} {\bibfnamefont {C.}~\bibnamefont {Sueur}},\ }\bibfield  {title}
  {\enquote {\bibinfo {title} {Collective decision making during group
  movements in {E}uropean bison, \emph{{B}ison bonasus}},}\ }\href@noop {}
  {\bibfield  {journal} {\bibinfo  {journal} {Animal Behaviour}\ }\textbf
  {\bibinfo {volume} {109}},\ \bibinfo {pages} {149--160} (\bibinfo {year}
  {2015})}\BibitemShut {NoStop}%
\bibitem [{\citenamefont {Merkle}, \citenamefont {Sigaud},\ and\ \citenamefont
  {Fortin}(2015)}]{merkle2015follow}%
  \BibitemOpen
  \bibfield  {author} {\bibinfo {author} {\bibfnamefont {J.~A.}\ \bibnamefont
  {Merkle}}, \bibinfo {author} {\bibfnamefont {M.}~\bibnamefont {Sigaud}}, \
  and\ \bibinfo {author} {\bibfnamefont {D.}~\bibnamefont {Fortin}},\
  }\bibfield  {title} {\enquote {\bibinfo {title} {To follow or not? {H}ow
  animals in fusion--fission societies handle conflicting information during
  group decision-making},}\ }\href@noop {} {\bibfield  {journal} {\bibinfo
  {journal} {Ecology Letters}\ } (\bibinfo {year} {2015})}\BibitemShut
  {NoStop}%
\bibitem [{\citenamefont {Schreier}\ and\ \citenamefont
  {Swedell}(2012)}]{schreier2012ecology}%
  \BibitemOpen
  \bibfield  {author} {\bibinfo {author} {\bibfnamefont {A.~L.}\ \bibnamefont
  {Schreier}}\ and\ \bibinfo {author} {\bibfnamefont {L.}~\bibnamefont
  {Swedell}},\ }\bibfield  {title} {\enquote {\bibinfo {title} {Ecology and
  sociality in a multilevel society: {E}cological determinants of spatial
  cohesion in hamadryas baboons},}\ }\href@noop {} {\bibfield  {journal}
  {\bibinfo  {journal} {American Journal of Physical Anthropology}\ }\textbf
  {\bibinfo {volume} {148}},\ \bibinfo {pages} {580--588} (\bibinfo {year}
  {2012})}\BibitemShut {NoStop}%
\bibitem [{\citenamefont {Ruckstuhl}(1998)}]{ruckstuhl1998foraging}%
  \BibitemOpen
  \bibfield  {author} {\bibinfo {author} {\bibfnamefont {K.~E.}\ \bibnamefont
  {Ruckstuhl}},\ }\bibfield  {title} {\enquote {\bibinfo {title} {Foraging
  behaviour and sexual segregation in bighorn sheep},}\ }\href@noop {}
  {\bibfield  {journal} {\bibinfo  {journal} {Animal Behaviour}\ }\textbf
  {\bibinfo {volume} {56}},\ \bibinfo {pages} {99--106} (\bibinfo {year}
  {1998})}\BibitemShut {NoStop}%
\bibitem [{\citenamefont {Conradt}\ and\ \citenamefont
  {Roper}(2003)}]{conradt2003group}%
  \BibitemOpen
  \bibfield  {author} {\bibinfo {author} {\bibfnamefont {L.}~\bibnamefont
  {Conradt}}\ and\ \bibinfo {author} {\bibfnamefont {T.~J.}\ \bibnamefont
  {Roper}},\ }\bibfield  {title} {\enquote {\bibinfo {title} {Group
  decision-making in animals},}\ }\href@noop {} {\bibfield  {journal} {\bibinfo
   {journal} {Nature}\ }\textbf {\bibinfo {volume} {421}},\ \bibinfo {pages}
  {155--158} (\bibinfo {year} {2003})}\BibitemShut {NoStop}%
\bibitem [{\citenamefont {King}\ and\ \citenamefont
  {Cowlishaw}(2009)}]{king2009leaders}%
  \BibitemOpen
  \bibfield  {author} {\bibinfo {author} {\bibfnamefont {A.~J.}\ \bibnamefont
  {King}}\ and\ \bibinfo {author} {\bibfnamefont {G.}~\bibnamefont
  {Cowlishaw}},\ }\bibfield  {title} {\enquote {\bibinfo {title} {Leaders,
  followers, and group decision-making},}\ }\href@noop {} {\bibfield  {journal}
  {\bibinfo  {journal} {Communicative \& Integrative biology}\ }\textbf
  {\bibinfo {volume} {2}},\ \bibinfo {pages} {147--150} (\bibinfo {year}
  {2009})}\BibitemShut {NoStop}%
\bibitem [{\citenamefont {Kerth}(2010)}]{kerth2010group}%
  \BibitemOpen
  \bibfield  {author} {\bibinfo {author} {\bibfnamefont {G.}~\bibnamefont
  {Kerth}},\ }\bibfield  {title} {\enquote {\bibinfo {title} {Group
  decision-making in animal societies},}\ }in\ \href@noop {} {\emph {\bibinfo
  {booktitle} {Animal Behaviour: Evolution and Mechanisms}}}\ (\bibinfo
  {publisher} {Springer},\ \bibinfo {year} {2010})\ pp.\ \bibinfo {pages}
  {241--265}\BibitemShut {NoStop}%
\bibitem [{\citenamefont {Krohn}\ and\ \citenamefont
  {Munksgaard}(1993)}]{krohn1993behaviour}%
  \BibitemOpen
  \bibfield  {author} {\bibinfo {author} {\bibfnamefont {C.~C.}\ \bibnamefont
  {Krohn}}\ and\ \bibinfo {author} {\bibfnamefont {L.}~\bibnamefont
  {Munksgaard}},\ }\bibfield  {title} {\enquote {\bibinfo {title} {Behaviour of
  dairy cows kept in extensive (loose housing/pasture) or intensive (tie stall)
  environments {II}. {L}ying and lying-down behaviour},}\ }\href@noop {}
  {\bibfield  {journal} {\bibinfo  {journal} {Applied Animal Behaviour
  Science}\ }\textbf {\bibinfo {volume} {37}},\ \bibinfo {pages} {1--16}
  (\bibinfo {year} {1993})}\BibitemShut {NoStop}%
\bibitem [{\citenamefont {Kilgour}(2012)}]{kilgour2012pursuit}%
  \BibitemOpen
  \bibfield  {author} {\bibinfo {author} {\bibfnamefont {R.~J.}\ \bibnamefont
  {Kilgour}},\ }\bibfield  {title} {\enquote {\bibinfo {title} {In pursuit of
  `normal': {A} review of the behaviour of cattle at pasture},}\ }\href@noop {}
  {\bibfield  {journal} {\bibinfo  {journal} {Applied Animal Behaviour
  Science}\ }\textbf {\bibinfo {volume} {138}},\ \bibinfo {pages} {1--11}
  (\bibinfo {year} {2012})}\BibitemShut {NoStop}%
\bibitem [{\citenamefont {Schirmann}\ \emph {et~al.}(2012)\citenamefont
  {Schirmann}, \citenamefont {Chapinal}, \citenamefont {Weary}, \citenamefont
  {Heuwieser},\ and\ \citenamefont
  {Von~Keyserlingk}}]{schirmann2012rumination}%
  \BibitemOpen
  \bibfield  {author} {\bibinfo {author} {\bibfnamefont {K.}~\bibnamefont
  {Schirmann}}, \bibinfo {author} {\bibfnamefont {N.}~\bibnamefont {Chapinal}},
  \bibinfo {author} {\bibfnamefont {D.~M.}\ \bibnamefont {Weary}}, \bibinfo
  {author} {\bibfnamefont {W.}~\bibnamefont {Heuwieser}}, \ and\ \bibinfo
  {author} {\bibfnamefont {M.~A.~G.}\ \bibnamefont {Von~Keyserlingk}},\
  }\bibfield  {title} {\enquote {\bibinfo {title} {Rumination and its
  relationship to feeding and lying behavior in {H}olstein dairy cows},}\
  }\href@noop {} {\bibfield  {journal} {\bibinfo  {journal} {Journal of Dairy
  Science}\ }\textbf {\bibinfo {volume} {95}},\ \bibinfo {pages} {3212--3217}
  (\bibinfo {year} {2012})}\BibitemShut {NoStop}%
\bibitem [{\citenamefont {Rob{\'e}rt}\ \emph {et~al.}(2011)\citenamefont
  {Rob{\'e}rt}, \citenamefont {White}, \citenamefont {Renter},\ and\
  \citenamefont {Larson}}]{robert2011determination}%
  \BibitemOpen
  \bibfield  {author} {\bibinfo {author} {\bibfnamefont {B.~D.}\ \bibnamefont
  {Rob{\'e}rt}}, \bibinfo {author} {\bibfnamefont {B.~J.}\ \bibnamefont
  {White}}, \bibinfo {author} {\bibfnamefont {D.~G.}\ \bibnamefont {Renter}}, \
  and\ \bibinfo {author} {\bibfnamefont {R.~L.}\ \bibnamefont {Larson}},\
  }\bibfield  {title} {\enquote {\bibinfo {title} {Determination of lying
  behavior patterns in healthy beef cattle by use of wireless
  accelerometers},}\ }\href@noop {} {\bibfield  {journal} {\bibinfo  {journal}
  {American Journal of Veterinary Research}\ }\textbf {\bibinfo {volume}
  {72}},\ \bibinfo {pages} {467--473} (\bibinfo {year} {2011})}\BibitemShut
  {NoStop}%
\bibitem [{\citenamefont {Mogensen}\ \emph {et~al.}(1997)\citenamefont
  {Mogensen}, \citenamefont {Krohn}, \citenamefont {S{\o}rensen}, \citenamefont
  {Hindhede},\ and\ \citenamefont {Nielsen}}]{mogensen1997association}%
  \BibitemOpen
  \bibfield  {author} {\bibinfo {author} {\bibfnamefont {L.}~\bibnamefont
  {Mogensen}}, \bibinfo {author} {\bibfnamefont {C.~C.}\ \bibnamefont {Krohn}},
  \bibinfo {author} {\bibfnamefont {J.~T.}\ \bibnamefont {S{\o}rensen}},
  \bibinfo {author} {\bibfnamefont {J.}~\bibnamefont {Hindhede}}, \ and\
  \bibinfo {author} {\bibfnamefont {L.~H.}\ \bibnamefont {Nielsen}},\
  }\bibfield  {title} {\enquote {\bibinfo {title} {Association between resting
  behaviour and live weight gain in dairy heifers housed in pens with different
  space allowance and floor type},}\ }\href@noop {} {\bibfield  {journal}
  {\bibinfo  {journal} {Applied Animal Behaviour Science}\ }\textbf {\bibinfo
  {volume} {55}},\ \bibinfo {pages} {11--19} (\bibinfo {year}
  {1997})}\BibitemShut {NoStop}%
\bibitem [{\citenamefont {Nielsen}\ \emph {et~al.}(1997)\citenamefont
  {Nielsen}, \citenamefont {Mogensen}, \citenamefont {Krohn}, \citenamefont
  {Hindhede},\ and\ \citenamefont {S{\o}rensen}}]{nielsen1997resting}%
  \BibitemOpen
  \bibfield  {author} {\bibinfo {author} {\bibfnamefont {L.~H.}\ \bibnamefont
  {Nielsen}}, \bibinfo {author} {\bibfnamefont {L.}~\bibnamefont {Mogensen}},
  \bibinfo {author} {\bibfnamefont {C.}~\bibnamefont {Krohn}}, \bibinfo
  {author} {\bibfnamefont {J.}~\bibnamefont {Hindhede}}, \ and\ \bibinfo
  {author} {\bibfnamefont {J.~T.}\ \bibnamefont {S{\o}rensen}},\ }\bibfield
  {title} {\enquote {\bibinfo {title} {Resting and social behaviour of dairy
  heifers housed in slatted floor pens with different sized bedded lying
  areas},}\ }\href@noop {} {\bibfield  {journal} {\bibinfo  {journal} {Applied
  Animal Behaviour Science}\ }\textbf {\bibinfo {volume} {54}},\ \bibinfo
  {pages} {307--316} (\bibinfo {year} {1997})}\BibitemShut {NoStop}%
\bibitem [{\citenamefont {Fisher}\ \emph {et~al.}(2002)\citenamefont {Fisher},
  \citenamefont {Verkerk}, \citenamefont {Morrow},\ and\ \citenamefont
  {Matthews}}]{fisher2002effects}%
  \BibitemOpen
  \bibfield  {author} {\bibinfo {author} {\bibfnamefont {A.~D.}\ \bibnamefont
  {Fisher}}, \bibinfo {author} {\bibfnamefont {G.~A.}\ \bibnamefont {Verkerk}},
  \bibinfo {author} {\bibfnamefont {C.~J.}\ \bibnamefont {Morrow}}, \ and\
  \bibinfo {author} {\bibfnamefont {L.~R.}\ \bibnamefont {Matthews}},\
  }\bibfield  {title} {\enquote {\bibinfo {title} {The effects of feed
  restriction and lying deprivation on pituitary--adrenal axis regulation in
  lactating cows},}\ }\href@noop {} {\bibfield  {journal} {\bibinfo  {journal}
  {Livestock Production Science}\ }\textbf {\bibinfo {volume} {73}},\ \bibinfo
  {pages} {255--263} (\bibinfo {year} {2002})}\BibitemShut {NoStop}%
\bibitem [{\citenamefont {Munksgaard}\ \emph {et~al.}(2005)\citenamefont
  {Munksgaard}, \citenamefont {Jensen}, \citenamefont {Pedersen}, \citenamefont
  {Hansen},\ and\ \citenamefont {Matthews}}]{munksgaard2005quantifying}%
  \BibitemOpen
  \bibfield  {author} {\bibinfo {author} {\bibfnamefont {L.}~\bibnamefont
  {Munksgaard}}, \bibinfo {author} {\bibfnamefont {M.~B.}\ \bibnamefont
  {Jensen}}, \bibinfo {author} {\bibfnamefont {L.~J.}\ \bibnamefont
  {Pedersen}}, \bibinfo {author} {\bibfnamefont {S.~W.}\ \bibnamefont
  {Hansen}}, \ and\ \bibinfo {author} {\bibfnamefont {L.}~\bibnamefont
  {Matthews}},\ }\bibfield  {title} {\enquote {\bibinfo {title} {Quantifying
  behavioural priorities --- {E}ffects of time constraints on behaviour of
  dairy cows, \emph{Bos taurus}},}\ }\href@noop {} {\bibfield  {journal}
  {\bibinfo  {journal} {Applied Animal Behaviour Science}\ }\textbf {\bibinfo
  {volume} {92}},\ \bibinfo {pages} {3--14} (\bibinfo {year}
  {2005})}\BibitemShut {NoStop}%
\bibitem [{\citenamefont {Albon}, \citenamefont {Clutton-Brock},\ and\
  \citenamefont {Langvatn}(1992)}]{albon1992cohort}%
  \BibitemOpen
  \bibfield  {author} {\bibinfo {author} {\bibfnamefont {S.~D.}\ \bibnamefont
  {Albon}}, \bibinfo {author} {\bibfnamefont {T.~H.}\ \bibnamefont
  {Clutton-Brock}}, \ and\ \bibinfo {author} {\bibfnamefont {R.}~\bibnamefont
  {Langvatn}},\ }\bibfield  {title} {\enquote {\bibinfo {title} {Cohort
  variation in reproduction and survival: Implications for population
  demography},}\ }in\ \href@noop {} {\emph {\bibinfo {booktitle} {The Biology
  of Deer}}}\ (\bibinfo  {publisher} {Springer},\ \bibinfo {year} {1992})\ pp.\
  \bibinfo {pages} {15--21}\BibitemShut {NoStop}%
\bibitem [{\citenamefont {{Shankar Raman}}(1997)}]{raman1997factors}%
  \BibitemOpen
  \bibfield  {author} {\bibinfo {author} {\bibfnamefont {T.~R.}\ \bibnamefont
  {{Shankar Raman}}},\ }\bibfield  {title} {\enquote {\bibinfo {title} {Factors
  influencing seasonal and monthly changes in the group size of chital or axis
  deer in southern {I}ndia},}\ }\href@noop {} {\bibfield  {journal} {\bibinfo
  {journal} {Journal of Biosciences}\ }\textbf {\bibinfo {volume} {22}},\
  \bibinfo {pages} {203--218} (\bibinfo {year} {1997})}\BibitemShut {NoStop}%
\bibitem [{\citenamefont {Clutton-Brock}, \citenamefont {Guinness},\ and\
  \citenamefont {Albon}(1982)}]{clutton1982red}%
  \BibitemOpen
  \bibfield  {author} {\bibinfo {author} {\bibfnamefont {T.~H.}\ \bibnamefont
  {Clutton-Brock}}, \bibinfo {author} {\bibfnamefont {F.~E.}\ \bibnamefont
  {Guinness}}, \ and\ \bibinfo {author} {\bibfnamefont {S.~D.}\ \bibnamefont
  {Albon}},\ }\href@noop {} {\emph {\bibinfo {title} {Red Deer: {B}ehavior and
  Ecology of Two Sexes}}}\ (\bibinfo  {publisher} {University of Chicago
  Press},\ \bibinfo {year} {1982})\BibitemShut {NoStop}%
\bibitem [{\citenamefont {Pulliam}(1973)}]{pulliam1973advantages}%
  \BibitemOpen
  \bibfield  {author} {\bibinfo {author} {\bibfnamefont {H.~R.}\ \bibnamefont
  {Pulliam}},\ }\bibfield  {title} {\enquote {\bibinfo {title} {On the
  advantages of flocking},}\ }\href@noop {} {\bibfield  {journal} {\bibinfo
  {journal} {Journal of theoretical Biology}\ }\textbf {\bibinfo {volume}
  {38}},\ \bibinfo {pages} {419--422} (\bibinfo {year} {1973})}\BibitemShut
  {NoStop}%
\bibitem [{\citenamefont {Foster}\ and\ \citenamefont
  {Treherne}(1981)}]{foster1981evidence}%
  \BibitemOpen
  \bibfield  {author} {\bibinfo {author} {\bibfnamefont {W.}~\bibnamefont
  {Foster}}\ and\ \bibinfo {author} {\bibfnamefont {J.}~\bibnamefont
  {Treherne}},\ }\bibfield  {title} {\enquote {\bibinfo {title} {Evidence for
  the dilution effect in the selfish herd from fish predation on a marine
  insect},}\ }\href@noop {} {\  (\bibinfo {year} {1981})}\BibitemShut {NoStop}%
\bibitem [{\citenamefont {Davies}, \citenamefont {Krebs},\ and\ \citenamefont
  {West}(2012)}]{davies2012introduction}%
  \BibitemOpen
  \bibfield  {author} {\bibinfo {author} {\bibfnamefont {N.~B.}\ \bibnamefont
  {Davies}}, \bibinfo {author} {\bibfnamefont {J.~R.}\ \bibnamefont {Krebs}}, \
  and\ \bibinfo {author} {\bibfnamefont {S.~A.}\ \bibnamefont {West}},\
  }\href@noop {} {\emph {\bibinfo {title} {An introduction to behavioural
  ecology}}}\ (\bibinfo  {publisher} {John Wiley \& Sons},\ \bibinfo {year}
  {2012})\BibitemShut {NoStop}%
\bibitem [{\citenamefont {Neill}\ and\ \citenamefont
  {Cullen}(1974)}]{neill1974experiments}%
  \BibitemOpen
  \bibfield  {author} {\bibinfo {author} {\bibfnamefont {S.}~\bibnamefont
  {Neill}}\ and\ \bibinfo {author} {\bibfnamefont {J.}~\bibnamefont {Cullen}},\
  }\bibfield  {title} {\enquote {\bibinfo {title} {Experiments on whether
  schooling by their prey affects the hunting behaviour of cephalopods and fish
  predators},}\ }\href@noop {} {\bibfield  {journal} {\bibinfo  {journal}
  {Journal of Zoology}\ }\textbf {\bibinfo {volume} {172}},\ \bibinfo {pages}
  {549--569} (\bibinfo {year} {1974})}\BibitemShut {NoStop}%
\bibitem [{\citenamefont {Parrish}(1993)}]{parrish1993comparison}%
  \BibitemOpen
  \bibfield  {author} {\bibinfo {author} {\bibfnamefont {J.~K.}\ \bibnamefont
  {Parrish}},\ }\bibfield  {title} {\enquote {\bibinfo {title} {Comparison of
  the hunting behavior of four piscine predators attacking schooling prey},}\
  }\href@noop {} {\bibfield  {journal} {\bibinfo  {journal} {Ethology}\
  }\textbf {\bibinfo {volume} {95}},\ \bibinfo {pages} {233--246} (\bibinfo
  {year} {1993})}\BibitemShut {NoStop}%
\bibitem [{\citenamefont {Cresswell}\ and\ \citenamefont
  {Quinn}(2011)}]{cresswell2011predicting}%
  \BibitemOpen
  \bibfield  {author} {\bibinfo {author} {\bibfnamefont {W.}~\bibnamefont
  {Cresswell}}\ and\ \bibinfo {author} {\bibfnamefont {J.~L.}\ \bibnamefont
  {Quinn}},\ }\bibfield  {title} {\enquote {\bibinfo {title} {Predicting the
  optimal prey group size from predator hunting behaviour},}\ }\href@noop {}
  {\bibfield  {journal} {\bibinfo  {journal} {Journal of Animal Ecology}\
  }\textbf {\bibinfo {volume} {80}},\ \bibinfo {pages} {310--319} (\bibinfo
  {year} {2011})}\BibitemShut {NoStop}%
\bibitem [{\citenamefont {Elgar}\ and\ \citenamefont
  {Catterall}(1981)}]{elgar1981flocking}%
  \BibitemOpen
  \bibfield  {author} {\bibinfo {author} {\bibfnamefont {M.~A.}\ \bibnamefont
  {Elgar}}\ and\ \bibinfo {author} {\bibfnamefont {C.~P.}\ \bibnamefont
  {Catterall}},\ }\bibfield  {title} {\enquote {\bibinfo {title} {Flocking and
  predator surveillance in house sparrows: Test of an hypothesis},}\
  }\href@noop {} {\bibfield  {journal} {\bibinfo  {journal} {Animal behaviour}\
  }\textbf {\bibinfo {volume} {29}},\ \bibinfo {pages} {868--872} (\bibinfo
  {year} {1981})}\BibitemShut {NoStop}%
\bibitem [{\citenamefont {Krakauer}(1995)}]{krakauer1995groups}%
  \BibitemOpen
  \bibfield  {author} {\bibinfo {author} {\bibfnamefont {D.~C.}\ \bibnamefont
  {Krakauer}},\ }\bibfield  {title} {\enquote {\bibinfo {title} {Groups confuse
  predators by exploiting perceptual bottlenecks: a connectionist model of the
  confusion effect},}\ }\href@noop {} {\bibfield  {journal} {\bibinfo
  {journal} {Behavioral Ecology and Sociobiology}\ }\textbf {\bibinfo {volume}
  {36}},\ \bibinfo {pages} {421--429} (\bibinfo {year} {1995})}\BibitemShut
  {NoStop}%
\bibitem [{\citenamefont {Sun}\ \emph {et~al.}(2011)\citenamefont {Sun},
  \citenamefont {Bollt}, \citenamefont {Porter},\ and\ \citenamefont
  {Dawkins}}]{sun2011mathematical}%
  \BibitemOpen
  \bibfield  {author} {\bibinfo {author} {\bibfnamefont {J.}~\bibnamefont
  {Sun}}, \bibinfo {author} {\bibfnamefont {E.~M.}\ \bibnamefont {Bollt}},
  \bibinfo {author} {\bibfnamefont {M.~A.}\ \bibnamefont {Porter}}, \ and\
  \bibinfo {author} {\bibfnamefont {M.~S.}\ \bibnamefont {Dawkins}},\
  }\bibfield  {title} {\enquote {\bibinfo {title} {A mathematical model for the
  dynamics and synchronization of cows},}\ }\href@noop {} {\bibfield  {journal}
  {\bibinfo  {journal} {Physica D}\ }\textbf {\bibinfo {volume} {240}},\
  \bibinfo {pages} {1497--1509} (\bibinfo {year} {2011})}\BibitemShut {NoStop}%
\bibitem [{\citenamefont {Gouz{\'e}}\ and\ \citenamefont
  {Sari}(2002)}]{gouze2002class}%
  \BibitemOpen
  \bibfield  {author} {\bibinfo {author} {\bibfnamefont {J.-L.}\ \bibnamefont
  {Gouz{\'e}}}\ and\ \bibinfo {author} {\bibfnamefont {T.}~\bibnamefont
  {Sari}},\ }\bibfield  {title} {\enquote {\bibinfo {title} {A class of
  piecewise linear differential equations arising in biological models},}\
  }\href@noop {} {\bibfield  {journal} {\bibinfo  {journal} {Dynamical
  systems}\ }\textbf {\bibinfo {volume} {17}},\ \bibinfo {pages} {299--316}
  (\bibinfo {year} {2002})}\BibitemShut {NoStop}%
\bibitem [{\citenamefont {Stoye}, \citenamefont {Porter},\ and\ \citenamefont
  {Dawkins}(2012)}]{stoye2012synchronized}%
  \BibitemOpen
  \bibfield  {author} {\bibinfo {author} {\bibfnamefont {S.}~\bibnamefont
  {Stoye}}, \bibinfo {author} {\bibfnamefont {M.~A.}\ \bibnamefont {Porter}}, \
  and\ \bibinfo {author} {\bibfnamefont {M.~S.}\ \bibnamefont {Dawkins}},\
  }\bibfield  {title} {\enquote {\bibinfo {title} {Synchronized lying in cattle
  in relation to time of day},}\ }\href@noop {} {\bibfield  {journal} {\bibinfo
   {journal} {Livestock Science}\ }\textbf {\bibinfo {volume} {149}},\ \bibinfo
  {pages} {70--73} (\bibinfo {year} {2012})}\BibitemShut {NoStop}%
\bibitem [{\citenamefont {Frisch}\ and\ \citenamefont
  {Vercoe}(1977)}]{frisch1977food}%
  \BibitemOpen
  \bibfield  {author} {\bibinfo {author} {\bibfnamefont {J.~E.}\ \bibnamefont
  {Frisch}}\ and\ \bibinfo {author} {\bibfnamefont {J.~E.}\ \bibnamefont
  {Vercoe}},\ }\bibfield  {title} {\enquote {\bibinfo {title} {Food intake,
  eating rate, weight gains, metabolic rate and efficiency of feed utilization
  in \emph{{B}os taurus} and \emph{{B}os indicus} crossbred cattle},}\
  }\href@noop {} {\bibfield  {journal} {\bibinfo  {journal} {Animal
  Production}\ }\textbf {\bibinfo {volume} {25}},\ \bibinfo {pages} {343--358}
  (\bibinfo {year} {1977})}\BibitemShut {NoStop}%
\bibitem [{\citenamefont {Illius}\ and\ \citenamefont
  {Gordon}(1987)}]{illius1987allometry}%
  \BibitemOpen
  \bibfield  {author} {\bibinfo {author} {\bibfnamefont {A.~W.}\ \bibnamefont
  {Illius}}\ and\ \bibinfo {author} {\bibfnamefont {I.~J.}\ \bibnamefont
  {Gordon}},\ }\bibfield  {title} {\enquote {\bibinfo {title} {The allometry of
  food intake in grazing ruminants},}\ }\href@noop {} {\bibfield  {journal}
  {\bibinfo  {journal} {The Journal of Animal Ecology}\ ,\ \bibinfo {pages}
  {989--999}} (\bibinfo {year} {1987})}\BibitemShut {NoStop}%
\bibitem [{\citenamefont {Bode}\ \emph {et~al.}(2010)\citenamefont {Bode},
  \citenamefont {Faria}, \citenamefont {Franks}, \citenamefont {Krause},\ and\
  \citenamefont {Wood}}]{bode2010perceived}%
  \BibitemOpen
  \bibfield  {author} {\bibinfo {author} {\bibfnamefont {N.~W.}\ \bibnamefont
  {Bode}}, \bibinfo {author} {\bibfnamefont {J.~J.}\ \bibnamefont {Faria}},
  \bibinfo {author} {\bibfnamefont {D.~W.}\ \bibnamefont {Franks}}, \bibinfo
  {author} {\bibfnamefont {J.}~\bibnamefont {Krause}}, \ and\ \bibinfo {author}
  {\bibfnamefont {A.~J.}\ \bibnamefont {Wood}},\ }\bibfield  {title} {\enquote
  {\bibinfo {title} {How perceived threat increases synchronization in
  collectively moving animal groups},}\ }\href@noop {} {\bibfield  {journal}
  {\bibinfo  {journal} {Proceedings of the Royal Society of London B:
  Biological Sciences}\ ,\ \bibinfo {pages} {rspb20100855}} (\bibinfo {year}
  {2010})}\BibitemShut {NoStop}%
\bibitem [{\citenamefont {Estevez}, \citenamefont {Andersen},\ and\
  \citenamefont {N{\ae}vdal}(2007)}]{estevez2007group}%
  \BibitemOpen
  \bibfield  {author} {\bibinfo {author} {\bibfnamefont {I.}~\bibnamefont
  {Estevez}}, \bibinfo {author} {\bibfnamefont {I.-L.}\ \bibnamefont
  {Andersen}}, \ and\ \bibinfo {author} {\bibfnamefont {E.}~\bibnamefont
  {N{\ae}vdal}},\ }\bibfield  {title} {\enquote {\bibinfo {title} {Group size,
  density and social dynamics in farm animals},}\ }\href@noop {} {\bibfield
  {journal} {\bibinfo  {journal} {Applied Animal Behaviour Science}\ }\textbf
  {\bibinfo {volume} {103}},\ \bibinfo {pages} {185--204} (\bibinfo {year}
  {2007})}\BibitemShut {NoStop}%
\bibitem [{\citenamefont {Mendl}\ and\ \citenamefont
  {Held}(2001)}]{mendl2001living}%
  \BibitemOpen
  \bibfield  {author} {\bibinfo {author} {\bibfnamefont {M.}~\bibnamefont
  {Mendl}}\ and\ \bibinfo {author} {\bibfnamefont {S.}~\bibnamefont {Held}},\
  }\bibfield  {title} {\enquote {\bibinfo {title} {Living in groups: {A}n
  evolutionary perspective},}\ }\href@noop {} {\bibfield  {journal} {\bibinfo
  {journal} {Social Behaviour in Farm Animals. L. J. Keeling \& H. W. Gonyou
  (Eds.)}\ ,\ \bibinfo {pages} {7--36}} (\bibinfo {year} {2001})}\BibitemShut
  {NoStop}%
\bibitem [{\citenamefont {Martin}\ \emph {et~al.}(2011)\citenamefont {Martin},
  \citenamefont {Fabes}, \citenamefont {Hanish}, \citenamefont {Leonard},\ and\
  \citenamefont {Dinella}}]{martin2011experienced}%
  \BibitemOpen
  \bibfield  {author} {\bibinfo {author} {\bibfnamefont {C.~L.}\ \bibnamefont
  {Martin}}, \bibinfo {author} {\bibfnamefont {R.~A.}\ \bibnamefont {Fabes}},
  \bibinfo {author} {\bibfnamefont {L.}~\bibnamefont {Hanish}}, \bibinfo
  {author} {\bibfnamefont {S.}~\bibnamefont {Leonard}}, \ and\ \bibinfo
  {author} {\bibfnamefont {L.~M.}\ \bibnamefont {Dinella}},\ }\bibfield
  {title} {\enquote {\bibinfo {title} {Experienced and expected similarity to
  same-gender peers: Moving toward a comprehensive model of gender
  segregation},}\ }\href@noop {} {\bibfield  {journal} {\bibinfo  {journal}
  {Sex Roles}\ }\textbf {\bibinfo {volume} {65}},\ \bibinfo {pages} {421--434}
  (\bibinfo {year} {2011})}\BibitemShut {NoStop}%
\bibitem [{\citenamefont {Lloyd}(1982)}]{lloyd1982least}%
  \BibitemOpen
  \bibfield  {author} {\bibinfo {author} {\bibfnamefont {S.~P.}\ \bibnamefont
  {Lloyd}},\ }\bibfield  {title} {\enquote {\bibinfo {title} {Least squares
  quantization in pcm},}\ }\href@noop {} {\bibfield  {journal} {\bibinfo
  {journal} {IEEE Transactions on Information Theory}\ }\textbf {\bibinfo
  {volume} {28}},\ \bibinfo {pages} {129--137} (\bibinfo {year}
  {1982})}\BibitemShut {NoStop}%
\bibitem [{\citenamefont {Bertram}(1980)}]{bertram1980vigilance}%
  \BibitemOpen
  \bibfield  {author} {\bibinfo {author} {\bibfnamefont {B.~C.~R.}\
  \bibnamefont {Bertram}},\ }\bibfield  {title} {\enquote {\bibinfo {title}
  {Vigilance and group size in ostriches},}\ }\href@noop {} {\bibfield
  {journal} {\bibinfo  {journal} {Animal Behaviour}\ }\textbf {\bibinfo
  {volume} {28}},\ \bibinfo {pages} {278--286} (\bibinfo {year}
  {1980})}\BibitemShut {NoStop}%
\bibitem [{\citenamefont {Roberts}(1996)}]{roberts1996individual}%
  \BibitemOpen
  \bibfield  {author} {\bibinfo {author} {\bibfnamefont {G.}~\bibnamefont
  {Roberts}},\ }\bibfield  {title} {\enquote {\bibinfo {title} {Why individual
  vigilance declines as group size increases},}\ }\href@noop {} {\bibfield
  {journal} {\bibinfo  {journal} {Animal Behaviour}\ }\textbf {\bibinfo
  {volume} {51}},\ \bibinfo {pages} {1077--1086} (\bibinfo {year}
  {1996})}\BibitemShut {NoStop}%
\end{thebibliography}
%%%%%%%%%%%
\nocite{*}

\end{document}